\documentclass{SciPost}

\hypersetup{
    colorlinks,
    linkcolor={red!50!black},
    citecolor={blue!50!black},
    urlcolor={blue!80!black}
}

\usepackage[bitstream-charter]{mathdesign}
\usepackage{graphicx,color}
\usepackage{dcolumn}
\usepackage{braket}
\usepackage{mathtools}
\usepackage{ulem}
\usepackage{times}
\usepackage{physics}
\usepackage{latexsym}

\usepackage{amsmath, amssymb}
\usepackage{mathtools}
\usepackage{multirow}

\fancypagestyle{SPstyle}{
\fancyhf{}
\lhead{\colorbox{scipostblue}{\bf \color{white} ~SciPost Physics Core }}
\rhead{{\bf \color{scipostdeepblue} ~Submission }}

\fancyfoot[C]{\textbf{\thepage}}
}

\newcommand{\gx}{g^x}
\newcommand{\gy}{g^y}
\newcommand{\gz}{g^z}
\newcommand{\gi}{g^I}
\newcommand{\be}{\begin{eqnarray}}
\newcommand{\ee}{\end{eqnarray}}

\renewcommand{\theequation}{\arabic{equation}}

\begin{document}

\pagestyle{SPstyle}

\begin{center}{\Large \textbf{\color{scipostdeepblue}{
Continuous Reset-Induced Phase Transition in Measurement-Free Random Quantum Circuits\\
}}}\end{center}

\begin{center}\textbf{
Hinata Yokoyama\textsuperscript{1$\star$},
Kengo Anzai\textsuperscript{1},
Dina Syverud-Lindland\textsuperscript{2},
Yoshihito Kuno\textsuperscript{3}, and
Hiroaki Matsueda\textsuperscript{1,4}
}\end{center}

\begin{center}
{\bf 1} Department of Applied Physics, Graduate School of Engineering, Tohoku University, Sendai 980-8579, Japan
\\
{\bf 2} Department of Physics, Faculty of Natural Sciences, Norwegian University of Science and Technology, Høgskoleringen 5, 7491 Trondheim, Norway
\\
{\bf 3} Graduate School of Engineering Science, Akita University, Akita 010-8502, Japan
\\
{\bf 4} Center for Science and Innovation in Spintronics, Tohoku University, Sendai 980-8577, Japan
\\[\baselineskip]
$\star$ \href{mailto:email1}{\small yokoyama.hinata.p7@dc.tohoku.ac.jp}\
\end{center}

\section*{\color{scipostdeepblue}{Abstract}}
\textbf{\boldmath{%
We study a random unitary quantum circuit with only reset channels, which has high feasibility for real quantum devices. In particular, we investigate the many-body statistical physics properties, ``reset-induced'' entanglement phase transitions comparing the classical statistical picture in the large ``$d$'' limit of qudits. In the property of the reset-induced phase transition the parameter of qudit $d$ is essential. That is, the transition properties induced by the reset channel significantly depend on $d$. We numerically elucidate this statement employing efficient stabilizer circuit simulations for $d=2$. Specifically, critical fluctuations that grow with system size are observed near the transition point, and the finite-size scaling analyses yield data collapse consistent with a continuous mixed-state phase transition.}
This behavior differs from expectations based on the classical statistical mapping in the large-$d$ limit.
}

\vspace{\baselineskip}



\vspace{10pt}
\noindent\rule{\textwidth}{1pt}
\tableofcontents
\noindent\rule{\textwidth}{1pt}
\vspace{10pt}


\section{Introduction}
Over many decades, there has been vast progress in the experimental research of a quantum computer, a machine that would take advantage of the full complexity of a quantum wave function of many-particle systems to solve a computational problem\cite{Ladd2010QuantumComputers,martinis2015qubit,RIKEN2014QuantumDots,babbush2025grandchallengequantumapplications}. Quantum computers with 50-100 qubits may be able to outperform classical digital computers; however, noise in quantum circuits significantly limits its ability and applications to compute and carry out some efficient quantum algorithm for many practical problems \cite{Preskill2018quantumcomputingin}. From the point of view of noise, the study of many-body quantum dynamics in an open quantum system has been gaining increasing attention during the Noisy Intermediate Scale Quantum (NISQ) era. 

In an open quantum system, decoherence is unavoidable because of the interaction between the system and its surrounding environment. Consequently, the quantum state loses coherence.
In quest of non-trivial many-body states, conversely, decoherence does not necessarily sweep out non-trivial many-body pure states which have been extensively studied so far, but becomes an interesting source to create non-trivial mixed states in open quantum systems. 
For example, decoherence induces phase transitions in the sense of statistical mechanics\cite{Liu_2024}. Moreover, a quantum dephasing channel, a type of decoherence source, in the presence of some global symmetries drives a system towards a more mixed state and induces a strong-to-weak symmetry breaking transition\cite{PhysRevB.110.094106,PRXQuantum.6.010344}. This effect of the dephasing channel can be understood as an effective statistical spin model\cite{PhysRevLett.129.080501}. 
As an example, the dephasing channel plays a role in a symmetry breaking field in a monitored random unitary quantum circuit\cite{PhysRevB.108.L060302,PhysRevLett.129.080501}.
Therefore, understanding the role of decoherence is crucial for uncovering the interplay between dissipation, symmetry breaking, and emergent many-body behavior in open quantum systems.

Recently, another type of decoherence mechanism, known as the reset channel, has attracted attention. It is commonly used to initialize qubits for tasks in quantum computation and quantum communication. Moreover, it is highly feasible for real experiment circuits\cite{PhysRevLett.110.120501,Zhou_2021}. Compared with a dephasing channel, a reset channel forces the $i$-th qubit into the $\ket{0}$ state and removes its correlations with the rest of the system. 
It is unique in that, in contrast to other decoherence such as a dephasing channel, 
the reset channel purifies a system to a product state in the mixed state dynamics once it is applied to every single site.
Furthermore, in a random unitary quantum circuit under the reset channel, there exists a ``reset''-induced phase transition. In the presence of a reset probability in the bulk with $q=p/L$, where $p$ is the average number of reset channels applied ($0\le p\le L$) and $L$ is the system size, the entanglement phase transition occurs when the reset probability exceeds some critical value~\cite{Liu_2024}. 
In the previous study~\cite{Liu_2024}, this phase transition is expected to be first-order, arising from the competition between two distinct spin configurations after mapping the $(1+1)$-D circuit system with ``large-$d$ qudits'' onto $2$D effective statistical spin models. 
A related setup incorporating boundary reset effects has also been investigated in the context of the coding phase transition~\cite{PRXQuantum.5.030327}. It was shown that initially encoding Bell pairs via noiseless scrambling with a unitary circuit is sufficient to protect the information against noise of arbitrary strength until late times. These two phase transitions are detected using mutual information, which measures the correlation between two subsystems. In the large-$d$ limit, both phase transitions are expected to be first-order since the dominant two spin configurations discontinuously change as the dissipation strength increases~\cite{Liu_2024,PRXQuantum.5.030327}. For finite qudit dimensions, additional (thermal) fluctuations appear, but—as argued in Ref. \cite{PRXQuantum.5.030327}—these fluctuations do not alter the order of the transition. Consequently, the coding transitions remain first‑order for all qudit dimensions. 

However, 
several finite-dimensional aspects of the reset-induced transition remain unclear. While the effective classical spin mapping provides a useful large-$d$ description, its applicability to finite-dimensional qubit circuits is not obvious. In particular, it is not yet numerically established whether the transition in the ($d=2$) case is discontinuous or continuous.
Then, whether different physical observables exhibit distinct transition behaviors from a statistical-mechanics perspective, and how accurately the analytical classical spin mapping captures the entanglement with the environment and the associated reset-induced phase transition remain open questions. In this work, we address these questions.

To address these questions, we explore the properties of reset-only phase transitions by performing extensive numerical simulations ($d=2$ case) and mapping the system to an effective statistical spin model. 
The numerical simulation shows strong agreement with the analytical results derived from the statistical spin mapping, except for the region near the phase transition point due to the fluctuations arising from the degeneracy of the leading spin configurations. Importantly, we find that for the case $d=2$, the reset-induced phase transition is continuous, not first-order. We specifically show data collapse of entanglement-related-observables and variances in finite-size scaling analyses and the second derivatives of the observables. We also confirm that the fluctuations are seen in terms of variances as expected in Ref.\cite{PRXQuantum.5.030327}. These results support that the reset-induced phase transitions are continuous. 
Moreover, the extracted critical exponents are compared with those reported for related measurement-induced and noise-induced transitions, providing a quantitative description of the mixed-state transition.

Furthermore, this phase transition depends not only on the reset probability $q$ but also on the ratio of $T/L$. We also observe that the phase transitions dependent on the ratio of $T/L$ are continuous not appearing clear first order ones.


This paper is organized as follows. In Sec.~\ref{methodology}, we introduce the setup of a random unitary quantum circuit with only reset channels, including the details of the algorithm of the reset channel in terms of the stabilizer formalism. The observables measured in the numerical calculation are also explained. The behavior of $S_p$, as understood from a classical statical spin perspective, is elaborated in Sec.~\ref{domain_wall_picture}. The numerical results obtained using the stabilizer formalism are shown in Sec.~\ref{numerical_result}. We present the properties of the reset-induced phase transition, along with its time dependence. The effects of $d$ on the reset-induced phase transition, entanglement behavior, and the universality class are discussed in Sec.~\ref{discussion}. 
Sec.~\ref{conclusion} is devoted to the conclusion.

\section{Methodology}\label{methodology}
We explain the setup of quantum circuits used in this work. This work focuses on two-qubit random unitary gates and local reset channels. 
The reset channel can be efficiently manipulated in the stabilizer formalism \cite{Nielsen2011}. We shall explain the practically-numerical algorithm and observables measured in the numerical calculation.

\subsection{Setup of quantum circuit}
\begin{figure}[t]
\begin{center} 
\vspace{0.5cm}
\includegraphics[width=8.5cm]{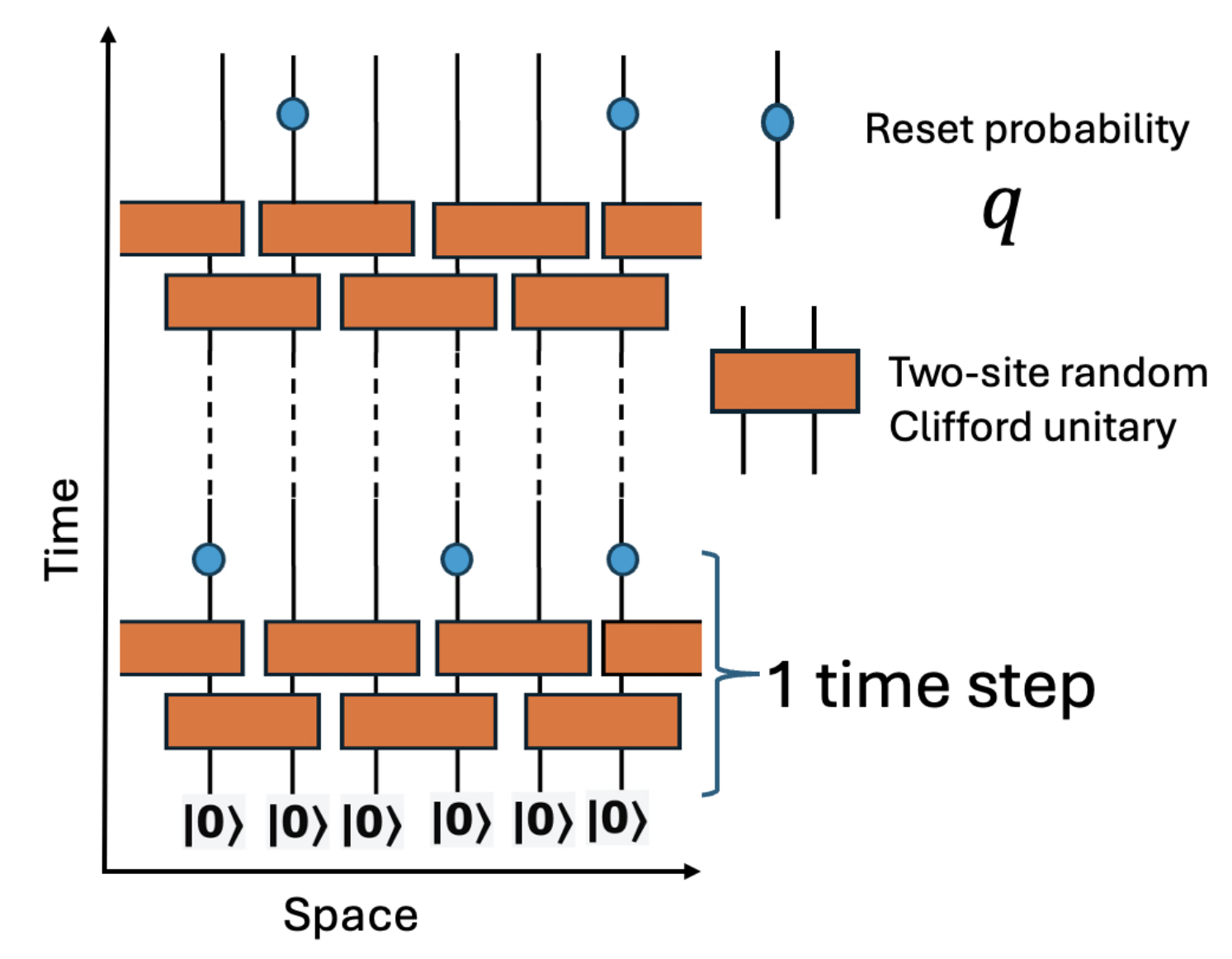}  
\end{center} 
\caption{A schematic representation of the random Clifford circuit subjected to the reset channel. The orange block represents a two-site random Clifford unitary gate, and the blue dot represents a reset channel. The horizontal and vertical axes represent the spatial and time directions, respectively. The reset channel is applied at each site with probability $q=p/L$. The system evolves from product states of $\ket{0}$ basis. Each time step includes two layer of the two-site unitary gates and the single layer reset channel with probability $q$ for each sites.}  
\label{reset_model}
\end{figure}

To investigate the reset-only dynamics of the random Clifford circuit, we consider a one-dimensional qubit chain with system size $L$ (corresponding to the number of qubits). We set an initial product state $\ket{0}^{\bigotimes L}$ and impose periodic boundary conditions (PBC). The circuit is illustrated as shown in Fig.~\ref{reset_model}. 
The circuit consists of two parts. The first is the two-site random Clifford unitary layers shown as the brickwork image in Fig.~\ref{reset_model}. Each two-site gate is randomly chosen from the Clifford group \cite{unitary_10.1063/1.4903507,pal_random2site_uni_PhysRevResearch.5.L012031}. The second is a reset-channel layer where the channel is locally applied on each site with size-dependent probability $q=p/L$. 
The unit of one time step in this work is shown in Fig.~\ref{reset_model}. A single time step consists of a unitary layer with random unitary gates on all even links, followed by a unitary layer with random unitary gates on all odd links, and a reset-channel layer. The total time step is $T$.
During each time step, we apply the layer of two-site random Clifford unitary gates and reset channels. 
Throughout this work, we set $T/L=\mathcal{O}(1)$.
Then, the parameter $p$ in the probability of the reset channel means that the total number of the application of the reset channel in $(1+1)$-D space-time plane. Indeed, we apply $p=\mathcal{O}(1)$ reset channels in the circuit plane, not depending on the size $L$. 

\subsection{Details of reset channel and its numerical algorithm}
The reset channel acting on a site $i$ is defined as:
\begin{equation}
R_i(\rho)=\mathrm{tr}_i(\rho)\otimes |0\rangle \langle0|_i,
\end{equation}
where $\rho$ is a density matrix and $i=1,\cdots,L$ is the site in which qubit initialization (reset) occurs\cite{Liu_2024,PhysRevLett.132.240402}.
This channel partially traces out over the degree of freedom on the site $i$ and inserts a qubit in a state of $|0\rangle \langle0|_i$ (see Appendix \ref{Appendix:A} for concrete examples). 
The initialization of qubits is often used before the start of computation in the context of quantum error correction. The simplest option to achieve this is to wait until the qubits decay to ground states \cite{geher2025resetresetquestion,DiVincenzo_2000}. However, the natural thermalization will be too slow for error correction purposes. Instead, cooling via projection will potentially lead to a shorter time scale than the natural relaxation time \cite{DiVincenzo_2000}. Furthermore, the reset channel purifies the system. Although by the combination with the random unitary gates the reset channel induces mixing (inducing entanglement with environment), it projects every qubit onto a product state (pure state) once the reset channel is applied to all the sites (An application example for two qubits is shown in Appendix $\ref{Appendix:A}$). 

We numerically implement this channel in the stabilizer formalism. Let us assume that we have a stabilizer group of $\mathcal{S}=\{g_l\}$ where $g_l$ is $k$ linearly-independent stabilizer generators with $1\le l\le k$.
Then, the density matrix of the system composed of $L$ qubits is given by $\rho=\frac{1}{2^{L-k}}\prod^k_{l=1}\frac{1+g_{l}}{2}$ \cite{PhysRevX.10.041020,PhysRevLett.132.240402}. For the state described by $\mathcal{S}$, we consider resetting a qubit at the $i$-th site. We define the number of stabilizer generators that include $X_i$, $Y_i,Z_i$, and $I_i$ as $n_x,n_y,n_z,n_I$ (satisfying $n_x+n_y+n_z+n_I=k\leq L$) and denote the corresponding stabilizer generators as $g^{x}_{1},\cdots,g^x_{n_x}$,$g^{y}_{1},\cdots,g^y_{n_y}$, $g^{z}_{1},\cdots,g^z_{n_z}$, and $g^{I}_{1},\cdots,g^I_{n_I}$ for $X_i,Y_i,Z_i,I_i$, respectively. 
Using the notation above, a set of stabilizer generators is expressed as \[\mathcal{S}=\{\gx_1,\cdots,\gx_{n_x},\gy_1,\cdots,\gy_{n_y},\gz_1,\cdots,\gz_{n_z}\}+\{\gi_1,\cdots,\gi_{n_I}\}.\] 
Here, in general, we operate the standard transformation on a given set of stabilizer generators \cite{Nielsen2011}.
We multiply the first generator with each other such that they have $I_i$. We obtain the different representation of a set of stabilizer generators, $\mathcal{S} \rightarrow \Tilde{\mathcal{S}}$:
\begin{align*}
    \Tilde{S}=&\{\gx_1,\gy_1,\gz_1\}+\{\gx_1\gx_2,\cdots,\gx_1\gx_{n_x},\gy_1\gy_2,\cdots,\gy_1\gy_{n_y},\gz_1\gz_2,\cdots,\gz_1\gz_{n_z},\gi_1,\cdots,\gi_{n_I}\},
\end{align*}
where $\rightarrow$ represents the manipulation of the stabilizer generators. 
Note that $g^\alpha_{n}g^{\alpha}_{m}=g^I=(\cdots)I_i(\cdots)$ with $\alpha=x,y,z$.

For this transformed set $\Tilde{\mathcal{S}}$, the reset channel $R_i(\rho)$ acts as (i) delete $\{\gx_1,\gy_1,\gz_1\}$ and add $\gx_1\gy_1\gz_1$, (ii) add $Z_i$ to $\Tilde{\mathcal{S}}$. The process (i) corresponds to $\mathrm{tr}_i(\rho)$ and the process (ii) corresponds to $\otimes |0\rangle\langle 0|$. 
Thus, the set of stabilizer generators subjected to $R_i$ becomes 
\begin{align*}
\Tilde{S}_{R_i}=&\{Z_i,\gx_1\gx_2,\cdots,\gx_1\gx_{n_x},\gy_1\gy_2,\cdots,\gy_1\gy_{n_y},\gz_1\gz_2,\cdots,\gz_1\gz_{n_z},\gi_1,\cdots,\gi_{n_I},\gx_1\gy_1\gz_1\}.
\end{align*}
Indeed, we obtain $\Tilde{\mathcal{S}}_{R_i}$ as a result of applying $ R_i(\rho)$.
In addition, there are rules for some special cases: 
\begin{enumerate}
    \item If $(n_x,n_y,n_z)$ is $(1,0,0),(0,1,0), (0,0,1),(1,1,0),$
    $(1,0,1),(0,1,1)$, we delete the stabilizer generator(s). 
    \item If $(n_x,n_y,n_z)$ is $(1,1,1)$, we multiply all the generators together and add them to the stabilizer group such that $\mathcal{S}=\{\gx_1,\gy_1,\gz_1\}+\{\gi_1,\cdots,\gi_{n_I}\}\rightarrow \ \{Z_i,\gx_1\gy_1\gz_1, \gi_1,\cdots,\gi_{n_I}\}$.
\end{enumerate}
In these rules, including the standard transformation, we essentially remove any stabilizers that contain $X_i,Y_i,Z_i$ by multiplication of stabilizer generators since $\mathrm{tr}_i[X_i]=\mathrm{tr}_i[Y_i]=\mathrm{tr}_i[Z_i]=0$. Then, by relabeling the set of the generator as $\Tilde{\mathcal{S}}_{R_i}=\{\tilde{g}_{l}\}$ with $l=1,2,\cdots, k'$
$R_i(\rho)=\frac{1}{2^{L-k}}\prod^{k'}_{l=1}\frac{1+\tilde{g}_{l}}{2}$ \cite{PhysRevX.10.041020,PhysRevLett.132.240402}.

This practical update way of the reset channel $R_i(\rho)$ based on the set of stabilizer generator is well-combined with the efficient numerical algorithm of the stabilizer table \cite{gottesman1998heisenbergrepresentationquantumcomputers,PhysRevA.70.052328} for the random Clifford unitary circuits. Thus, we can easily carry out numerical simulations up to large system sizes $L=\mathcal{O}(10^2)$.
\subsection{Observables}
In this work, we consider the following quantities, which are calculated using the efficient stabilizer algorithm\cite{gottesman1998heisenbergrepresentationquantumcomputers,PhysRevA.70.052328}. 

The first is logarithmic purity, denoted as $S_p$\cite{PhysRevX.10.041020}:
\begin{equation}
    S_p=-\text{log}_2\ \mathrm{tr}[\rho^2].
\end{equation}
It is also called second R\'{e}nyi entropy and measures the correlation between a system and an environment, namely system-environment entanglement\cite{Ashida2024}. In addition, $S_p$ measures the purity of the system in which $0\leq S_p/L\leq 1$. If $S_p/L=0$, the system is completely pure; otherwise, the system is mixed. 

The other important quantity is many-body negativity (MBN) defined as:
\begin{equation}
    E_N=\text{log}_2\ |\rho^{\Gamma_{A}}|_1,
\end{equation}
where $\Gamma_{A}$ is a partial transpose operation for the subsystem of $A=\frac{L}{2}$, and $|\cdot|_1$ is the trace norm. This quantity measures quantum entanglement in the system \cite{peres1996,Horodecki_family,plenio_nega_PhysRevLett.95.090503} and captures entanglement transitions in pure and mixed states\cite{PRXQuantum.2.030313,PhysRevLett.129.080501,sharma2022,Liu_2024,Anzai_2026}. 
The MBN is also computed in the stabilizer formalism:
\begin{equation}
    E_N=\frac{1}{2}{\rm rank}_{\mathbb{F}_2}(K^A),
\end{equation}
where $K^A$ is a matrix $m\times m$, and $m$ is a total number of independent stabilizer generators for a state of $\rho$ and $m\leq L$. The detailed derivation is given in \cite{PRXQuantum.2.030313,shi2021entanglementnegativitycriticalpoint}. 
Briefly, to construct the matrix $m\times m$ $K^A$, the total generator $m$-stabilizer is truncated within the subsystem $A$, and the truncated set is denoted by $\mathcal{S}^{tr}_A=\{ g^A_{l}\}$. Then, anti-commutative pairs of $g^A_{l}$'s appear. 
The presence of the pair means quantum correlations between the subsystem $A$ and the remaining part. 
By using $\mathcal{S}^{tr}_A=\{ g^A_{l}\}$, we construct the matrix $K^A$ given by \cite{PRXQuantum.2.030313}
\begin{eqnarray}
K^A_{l,l'}=
\begin{cases}
1 & \mbox{if}\:\: \{g^{A}_{l},g^{A}_{l'}\}=0\\
0 & \mbox{if}\:\: [ g^{A}_{l},g^{A}_{l'}]=0
\end{cases},
\label{K_rule}
\end{eqnarray}
where $\{g^{A}_{l},g^{A}_{l'}\}=0$ means that the truncated stabilizer generators $g^{A}_{\ell}$ and $g^{A}_{\ell'}$ are anti-commuting,
and $[g^{A}_{l},g^{A}_{l'}]=0$ means that the truncated stabilizer generators $g^{A}_{l}$ and $g^{A}_{l'}$ are commuting.
As a result, we obtain the matrix $K^A$ as the binary($Z_2$) $m \times m$ matrix and obtain the MBN.

In the later numerical calculations, we focus on the ensemble average for the two quantities $S_p$ and $E_N$. That is, we calculate many trajectory samples for the circuit dynamics. If each sample is labeled by $s$, we calculate the following quantities by using the stabilizer algorithm,
\begin{eqnarray}
\langle O\rangle=\mathbb{E}[O^s]=\frac{1}{N_s}\sum_{s}O^s,\nonumber
\end{eqnarray} 
where $O=S_p$ or $E_N$. $\langle\cdot\rangle$ represents the mean value, and $O^s$ represents a single sample obtained by the trajectory sample $s$. 

\section{Behavior of $S_p$ from a classical statistical mechanical spin picture}\label{domain_wall_picture}

Before we show the numerical calculations, we discuss the effective classical spin model, which is obtained from a mapping of the random unitary quantum circuit. Here, the mapping of the circuit subjected to the reset channel was first shown in \cite{Liu_2024}.

Based on the approach previously introduced in \cite{Liu_2024}, we develop a more precise formulation of the classical spin mapping and analytically investigate the qualitative behavior of $S_p$ in detail under several extreme assumptions. The precise formulation is summarized in Appendix \ref{Appendix:B}.

Using the mapping, the presence of the transition of $S_p$, especially for the large-$d$ qudit case, can be understood. 
We clearly expect the presence of a reset-induced phase transition and also estimate the dependence of $S_p$ on $p$ and $T$ in the large-$d$ limit, which is given by
\begin{eqnarray}
S_p \approx 
\begin{cases} 
Tp\log_2(d), & p \leq p_c, \\
S_0(p)L\log_2 (d), & p > p_c,\\ \end{cases}
\end{eqnarray}
where $p_c$ is the phase transition point if exists. This transition is induced by the change of the global spin configuration \cite{Liu_2024}, and for large-$d$, only a limited number of spin configurations remain dominant even around the transition point. Thus, in the large-$d$ regime, the reset-induced phase transition characterized by $S_p$ is the first-order phase transition. In other words, it is neither a continuous one, nor one exhibiting large fluctuations around the transition point. This expectation has been mentioned in \cite{Liu_2024}. 
However, we examine the result in detail.
In particular, we assess to what extent this large-$d$ result differs from, or agree with, the numerical results for $S_p$ obtained from the $d=2$ (qubit) stabilizer circuit.


\section{Numerical Results}\label{numerical_result}
In this section, we present numerical results to elucidate the properties of the reset-induced phase transitions and to support that the reset-induced phase transitions are continuous for the setup shown in Sec.~\ref{methodology}. 
We employ the stabilizer algorithm \cite{gottesman1998heisenbergrepresentationquantumcomputers,PhysRevA.70.052328}. The following subsections examine the transition for a fixed total time step $T=\mathcal{O}(L)$ and its $T/L$-dependence. 

\subsection{Reset-only dynamics at a fixed time}
\begin{figure}[t]
\begin{center} 
\vspace{0.5cm}
\includegraphics[width=8.5cm]{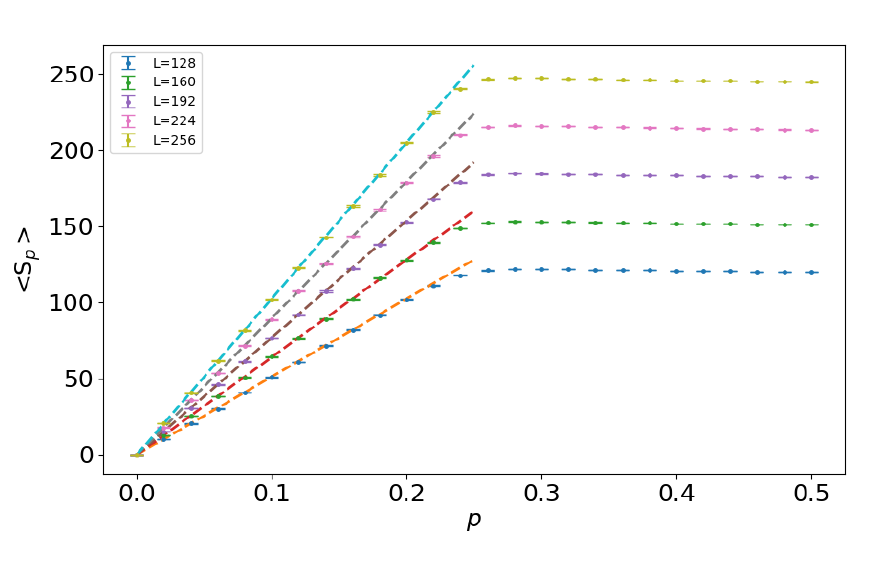} 
\end{center} 
\caption{(a) $p$-dependence of $S_p$. The data were taken from $L=128$ to $L=256$. 
The dashed lines represent analytical expectation of Eq.~(\ref{S_p_d=2_exp}) of $p<p_c$ case. For all data point, the trajectory samples are $\mathcal{O}(10^3)$ and the total time step is $T=4L$.}
\label{second_Renyi_entropy}
\end{figure}

First, we investigate the reset-induced phase transition under fixing $T/L=4$. We show the $p$-dependence of $S_p$ in the region $0\leq p\leq0.5$ where we see the change in $S_p$ very clearly. The results with different sizes are shown in Fig.~\ref{second_Renyi_entropy}.

We find that in the small-$p$ regime, $S_p$ increases linearly and also the values for different system size $L$ exhibit almost linear increase. This indicates that for small $p$, $S_p$ exhibits volume-law entanglement to environment. Then, for $p \gtrapprox 0.25$, $S_p$ saturates although the dependence on $L$ remains. Such a drastic change in the behavior $S_p$ around $p\sim 0.25$ indicates the presence of a phase transition, the details of which are discussed later.

This behavior is qualitatively captured by mapping the circuit to the effective spin model shown in Appendix \ref{Appendix:B}. As mentioned in Sec.~\ref{domain_wall_picture}, from the classical statistical spin configurations, we expect the presence of a phase transition, and the form of $S_p$ for large-$d$ is given. 
Here, we compare the numerical data with the analytical results, imposing $d=2$. Then, we have
\begin{eqnarray}\label{s_p_transition}
S_p \approx 
\begin{cases} 
TLq, & p \leq p_c, \\ 
S_0(p)L, & p > p_c,\\ 
\end{cases}
\label{S_p_d=2_exp}
\end{eqnarray}
where $S_0$ is $p$-dependent. 
We clearly find that the numerical data fairly correspond to $TLq=Tp$ in the small-$p$ regime, as shown in Fig.~\ref{second_Renyi_entropy}. This fact indicates that the analytical expectation for large-$d$ approximation in the mapping captures the $p$-dependence of $S_p$ in the small-$p$ regime even for shifting $d\to 2$ (changing from large-$d$ qudit to qubit).

We next discuss the comparison in the large-$p$ regime when $p \gtrapprox 0.25$. 
The numerical data for each $L$ exhibit a behavior similar to $S_0(p)L$. 
Thus, the phenomenological picture based on the classical spin configuration holds.
As explained in Appendix \ref{Appendix:B}, 
more specifically, the leading spin configuration are the domains of two permutation spins with the single domain wall. 
The one spin is 
$\mathbb{I}=
\begin{pmatrix}
1 & 2 & \cdots & n \\
1 & 2 & \cdots & n
\end{pmatrix}$ which is the identity permutation among the $n$-ket indices, this domain is due to the massive amount of ``noise'' (reset-channel). The other spin is 
$\mathbb{C}=\begin{pmatrix}
1 & 2 & \cdots & n \\
2 & 3 & \cdots & 1
\end{pmatrix}$, which is the cyclic permutation between the $n$ indices. The $\mathbb{C}$ domain is due to the fixed upper boundary.  
The appearance of a domain wall separates the $\mathbb{C}$ and $\mathbb{I}$ regions. 
In the large-$p$ regime, the logarithmic purity of our numerical results is equivalent to the form $S_p=S_0(p)L$, expected from the domain wall configuration.
Therefore, we expect that the behavior of $S_p$ in our numerics ($d\to 2$ case) is captured by the picture of the zig-zag crossing domain wall separating the two distinct domains.

\begin{table}[h]
\centering
\begin{tabular}{|*{13}{c|}}
\hline
$p$ & 0.28 & 0.32 & 0.36 & 0.40 & 0.44& 0.48  \\
\hline
$S_0(p)$  & 0.9819  &  0.9801 & 0.9779  & 0.9766  & 0.9759  & 0.9750   \\
\hline
\end{tabular}
\caption{The value of $S_0(p)$ corresponding to Fig.~\ref{second_Renyi_entropy}.}
\label{table_S_0}
\end{table}

We further estimate the numerical value of the tension $S_0$ of the domain wall from the numerical data in detail. 
We summarize $S_0 (p)$ relevant to Fig.~\ref{second_Renyi_entropy} for $p> p_c$ in Table~\ref{table_S_0}.
Regarding the estimated $S_0$, we find that the $p$-dependence is fairly small after the transition. However, as $p$ increases, $S_0$ decreases slightly. This means that the domain wall length in the effective spin picture decreases as more ``noise'' is introduced into the system.
Moreover, we comment on the extreme limits. In the case $p=L$ (i.e., $q=1$), we expect $S_0=S_p=0$. We interpret this as all states being projected onto product states. This interpretation is consistent with the numerical result shown in Fig.\ref{the_role_of_reset_channel} in Appendix \ref{Appendix:C}.

\begin{figure}[t]
\begin{center} 
\vspace{0.5cm}
\includegraphics[width=15cm]{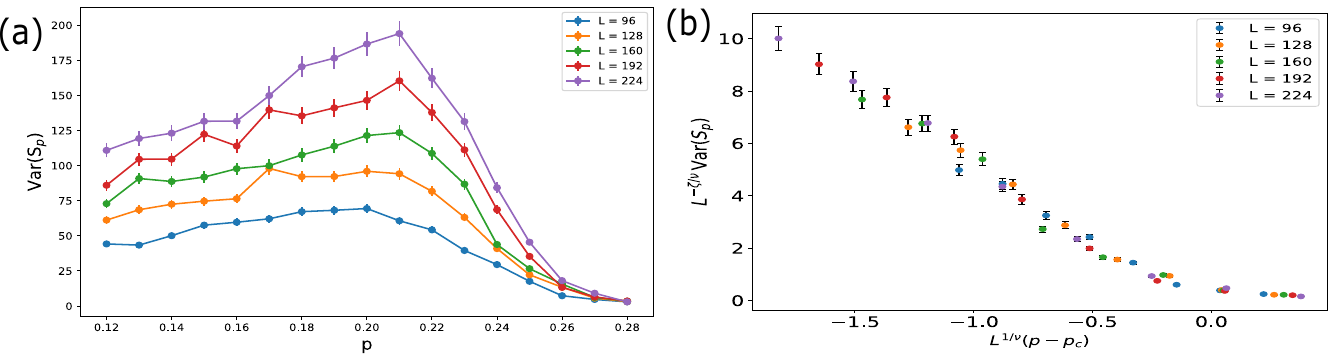}  
\end{center} 
\caption{(a)$p$-dependence of variance of $S_p$. The peak increases as $L$ is incremented. (b) The results of the FSS analysis at a point to the right of the peak. The extracted critical exponents and points are $\nu=1.57(22),\zeta=0.86(44)$, and $p_c=0.26$.}
\label{variance_purity_fss}
\end{figure}

We now discuss the behavior of $S_p$ around $p\sim 0.25$. There, the deviation between the theoretical prediction $TLq$ in Eq.~(\ref{s_p_transition}) and the numerical data becomes significant. This implies that 
in the spin configuration, small clusters of $\mathbb{I}$ bubbles emerge, leading to a breakdown of the large-$d$ approximation on the analytical side (see Appendix \ref{Appendix:B} for details) and the system fluctuation becomes large around the transition. 
Here, we show the variance obtained from the trajectory samples of $S_p$ in Fig.~\ref{variance_purity_fss}(a). We observe that the peak of the variance grows around $p\sim 0.21$ with an increase of $L$. This corresponds to the fluctuations seen in the finite qudit $d$ dimensions mentioned in Ref.\cite{PRXQuantum.5.030327}. 
We expect that in the thermodynamic limit, the variance tends to diverge. However, the peak position of variance of $S_p$, ${\rm Var}(S_p)$, in $p$ is smaller than the value of $p$ at which $S_p$ starts to saturate [See Fig.~\ref{second_Renyi_entropy}]. This deviation implies that, within our currently accessible computational resources, finite-size effects still remain, and the transition region is broadened rather than being resolved as a sharply defined single point in $p$. 
Consequently, although ${\rm Var}(S_p)$ alone does not allow a precise determination of the transition point or critical behavior, it still provides useful information about the characteristic features of the transition. We therefore perform a finite-size scaling (FSS) analysis of ${\rm Var}(S_p)$ using pyfssa\cite{melchert2009autoscalepyprogramautomatic}.  
We set the following ansatz
\begin{equation}\label{new_ansatz_S_p}
{\rm Var}(S_p(p))=L^\frac{\zeta}{\nu}f(L^{\frac{1}{\nu}}(p-p_c)),
\end{equation}
where $\nu$ is a correlation length critical exponent, $\zeta$ is a critical exponent and $p_c$ is the phase transition point~\cite{Anzai_2026,PhysRevLett.129.080501}.  
In the transition regime, we expect that the data collected from different system sizes collapse onto the same curve with the correct choice of $p_c$, $\nu$, and $\zeta$~\cite{Liu_2024,PRXQuantum.5.030327}. Here, allowing for a finite broadening of the transition region, we perform the FSS analysis on the right side of the peak, where the system-size dependence of $\mathrm{Var}(S_p)$ is gradually suppressed. We also perform the FSS analysis near the peak[See Appendix~\ref{variance_appendix}]. The former regime is close to the point where $S_p$ reaches the saturation regime~[See Fig.~\ref{second_Renyi_entropy}]. The result of the FSS analysis for the data on the right side of the peak is shown in Fig.~\ref{variance_purity_fss}(b). The extracted results of critical exponents and point are $\nu \simeq 1.6$, $\zeta\simeq 0.8$ and $p_c\simeq 0.26$. This extracted value of $p_c$ agrees with the point at which the scaling of $S_p$ is changed shown in Fig.\ref{second_Renyi_entropy}. 

\begin{figure}[t]
\begin{center} 
\vspace{0.5cm}
\includegraphics[width=13cm]{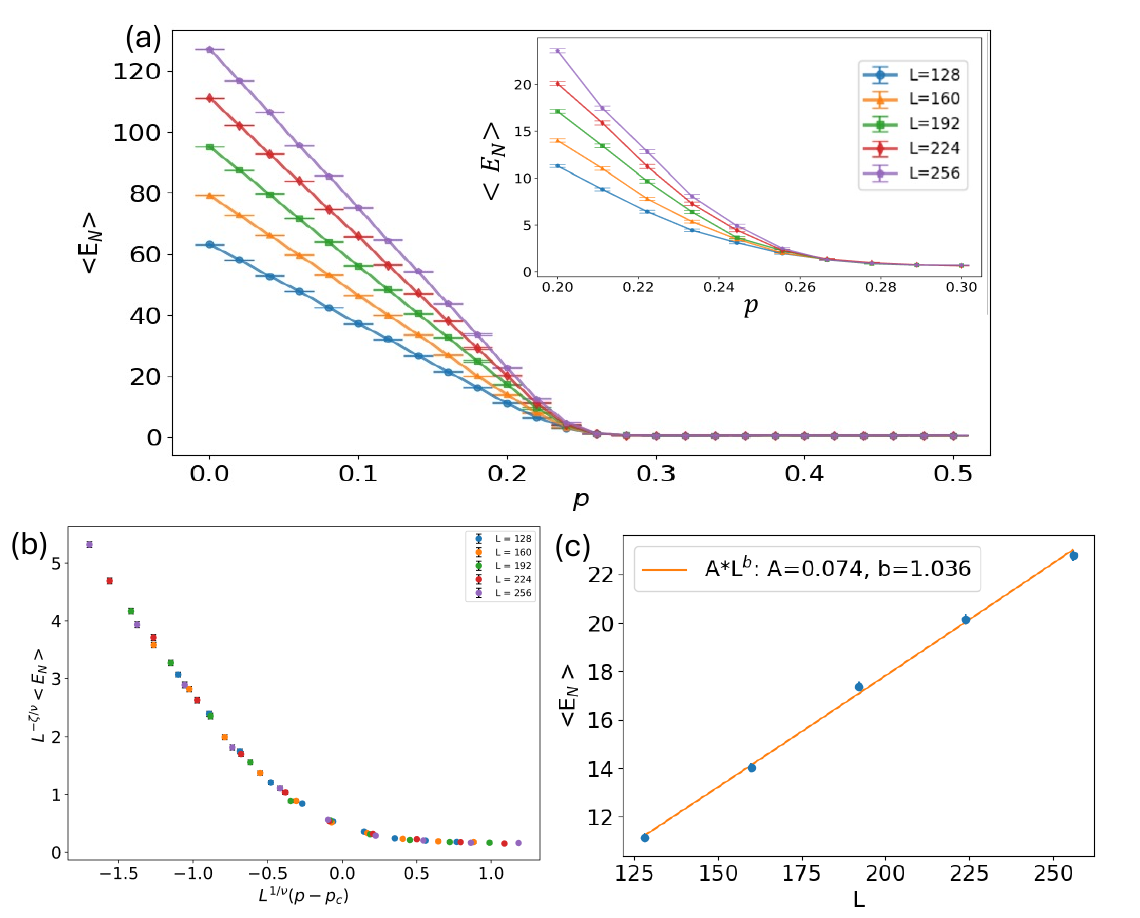}  
\end{center} 
\caption{(a) $p$-dependence of the MBN. The lines for different system sizes from $L=128$ to $L=256$ merge all together after $p\approx0.26$ (inset). (b) The results of the FSS analysis for the MBN. The numerically extracted results are $p_c=0.25$, $\zeta=0.43(18)$ and $\nu=1.61(31)$. (c) $L$-dependence of $E_N$ and the linear fitting for $p=0.20$. The data were averaged over $\mathcal{O}(10^3)$ at $T=4L$.}  
\label{reset_induced_phase_transition}
\end{figure}
Next, under the same setting, we show $p$-dependence of the MBN as shown in Fig.~\ref{reset_induced_phase_transition}(a).
As $p$ increases, the MBN for different system sizes starts to decrease monotonically. The data with different system size merges around $p=0.26$; that is, the system-size dependence vanishes, indicating an area-law phase. 
This implies the presence of the reset-induced phase transition.
To verify the presence of the phase transition, we again conduct the FSS analysis for the bare value of MBN by using the same ansatz in Eq.~(\ref{new_ansatz_S_p}).
The results of the FSS are shown in Fig.~\ref{reset_induced_phase_transition}(b). We observe the data collapse and estimate $p_c\simeq0.25$, 
$\nu=1.61\pm0.31$ and $\zeta=0.43\pm0.18$. These estimates are reasonably close to the scaling form used in Ref.~\cite{Liu_2024}, where $\nu=2$ and $\zeta=2/3$ were assumed for the collapse. Thus, despite the remaining uncertainty due to finite-size effects and the fitting window, our independent analysis supports the presence of a consistent scaling regime in the logarithmic negativity.

In addition, the transition separates the scaling law of entanglement. Before the transition (small-$p$), we observe a typical $L$-dependence of $E_N$ as shown in Fig.~\ref{reset_induced_phase_transition}(c). There, the linear scaling behavior in MBN appears near the critical point at $p=0.20$. The estimated exponent in the volume-law phase is $b\approx1$ where $E_N=L^{b}$. 
Conversely, when the MBN converges to a finite value for $p\geq p_c$, the area-law phase appears. 

\begin{figure}[t]
\begin{center} 
\vspace{0.5cm}
\includegraphics[width=15cm]{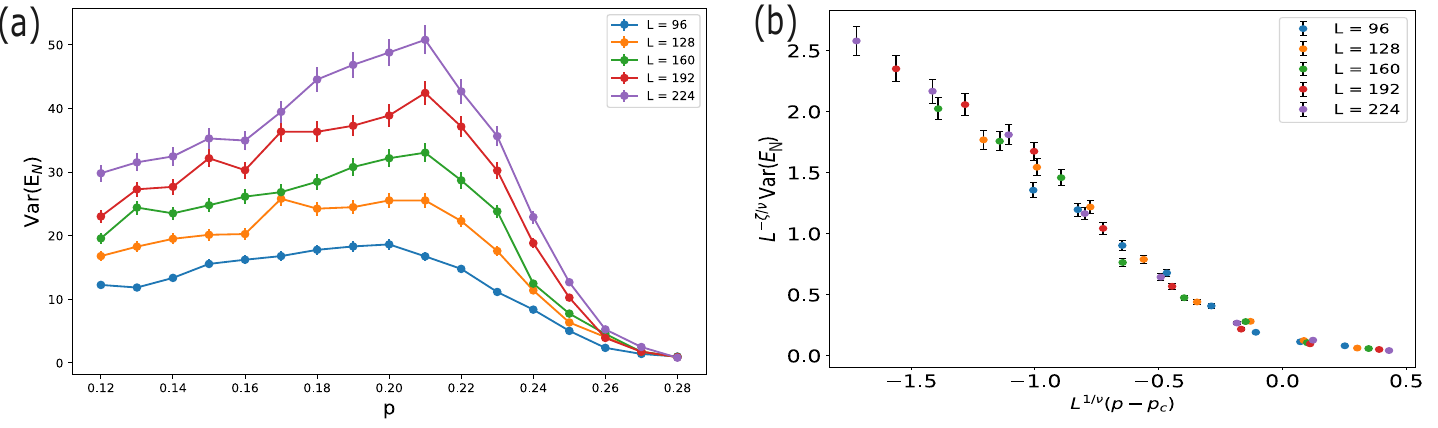}  
\end{center} 
\caption{(a) $p$-dependence of variance of $E_N$. The peak increases as $L$ is incremented. (b) The results of the FSS analysis at a point to the right of the peak. The extracted critical exponents and points are $\nu=1.58(27),\zeta=0.87(42)$, and $p_c=0.26(1)$.}
\label{variance_E_p}
\end{figure}
Moreover, we investigate the transition using the variance of the MBN. Figure~\ref{variance_E_p} shows the variance of $E_N$ and the result of its FSS analyzes. The peaks increase with an increase of $L$ around $p=0.21$ shown in Fig.~\ref{variance_E_p}(a). Analogously to the variance of the logarithmic purity, the finite size effects are still observed, which leads to broadening the transition region. 
Using the same ansatz in Eq.~(\ref{new_ansatz_S_p}), we perform the FSS analysis around a point to the right of the peak in Fig.~\ref{variance_E_p}(b). We observe data collapse where we estimate $\nu\simeq1.6$, $\zeta\simeq0.8$, and  $p_c\simeq0.26$.

\begin{figure}[t]
\begin{center} 
\vspace{0.5cm}
\includegraphics[width=15cm]{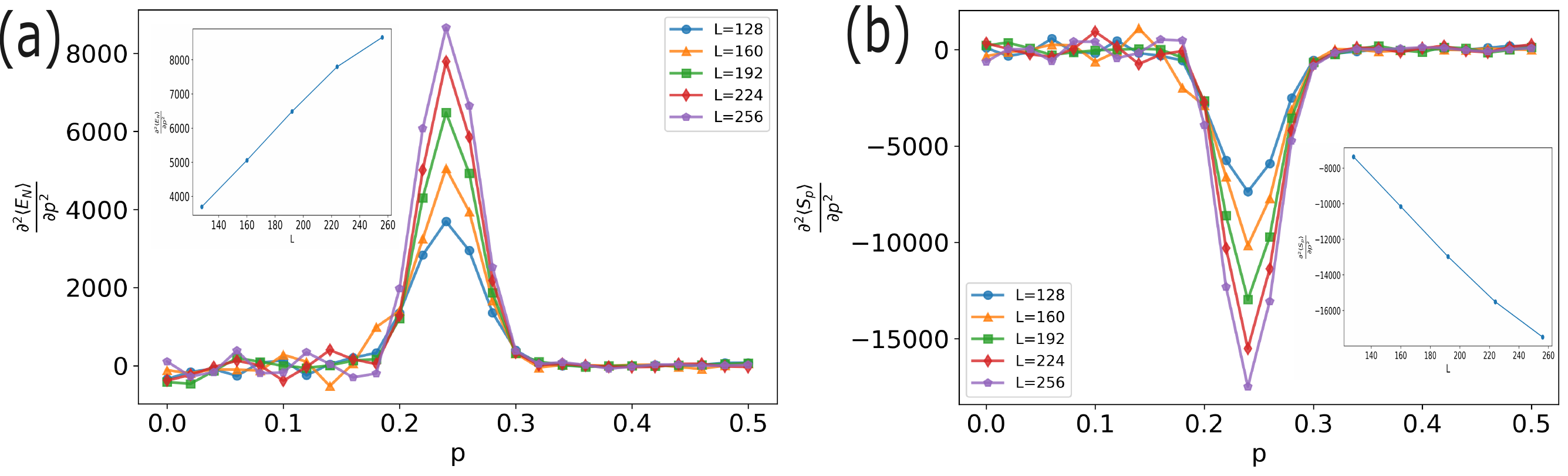}  
\end{center} 
\caption{$p$-dependence of the second derivative for $E_N$ [(a)] and $S_p$ [(b)], respectively. Both peaks are observed with $p=0.24$. Insets: the peak values grow linearly in system size. The absolute values of the peaks get large as increasing the system size $L$.}
\label{second_order_phase_transition}
\end{figure}

Importantly, as suggested in \cite{Liu_2024,PRXQuantum.5.030327}, the reset-induced phase transition (especially observed by $S_p$) is expected to be first-order. However, the numerical evidence, especially for the case $d=2$, has not been presented. The numerical results presented in this section already demonstrate the data collapse of the different observables. These well data collapses are consistent with a continuous transition. 
We further give the supporting data: $p$-dependence of the second derivative of both quantities of $E_p$ and $S_p$ as shown in Figs.~\ref{second_order_phase_transition}(a) and \ref{second_order_phase_transition}(b). The quantities tend to diverge as the system size increases near the critical point $p_c\sim 0.25$ as shown in each inset. This indicates that fluctuations seen in the finite qudit $d$-dimensions non-trivially affect the reset-induced phase transitions and can make them continuous. Therefore, we conclude that the reset-induced phase transitions characterized by $S_p$ and $E_N$ including the variances are continuous with large fluctuations near the critical point.

\subsection{$T/L$-dependence of reset-induced phase transition}
In the previous subsection, we examined the dynamics in a fixed time $T=4L$. The reset-induced phase transition is observed by controlling the parameter $p$. Furthermore, previous work (\cite{Liu_2024,PRXQuantum.5.030327}) suggests that this phase transition is also controlled by the total time step, $T$. Natural questions arise: (i) Whether or not the critical point of the reset-induced phase transition depends on $T$ and (ii) How robust the critical properties observed in the previous subsection are. 
In this subsection, we investigate the reset-induced phase transitions by varying the ratio $T/L$.

\begin{figure}[t]
\begin{center} 
\vspace{0.5cm}
\includegraphics[width=7cm]{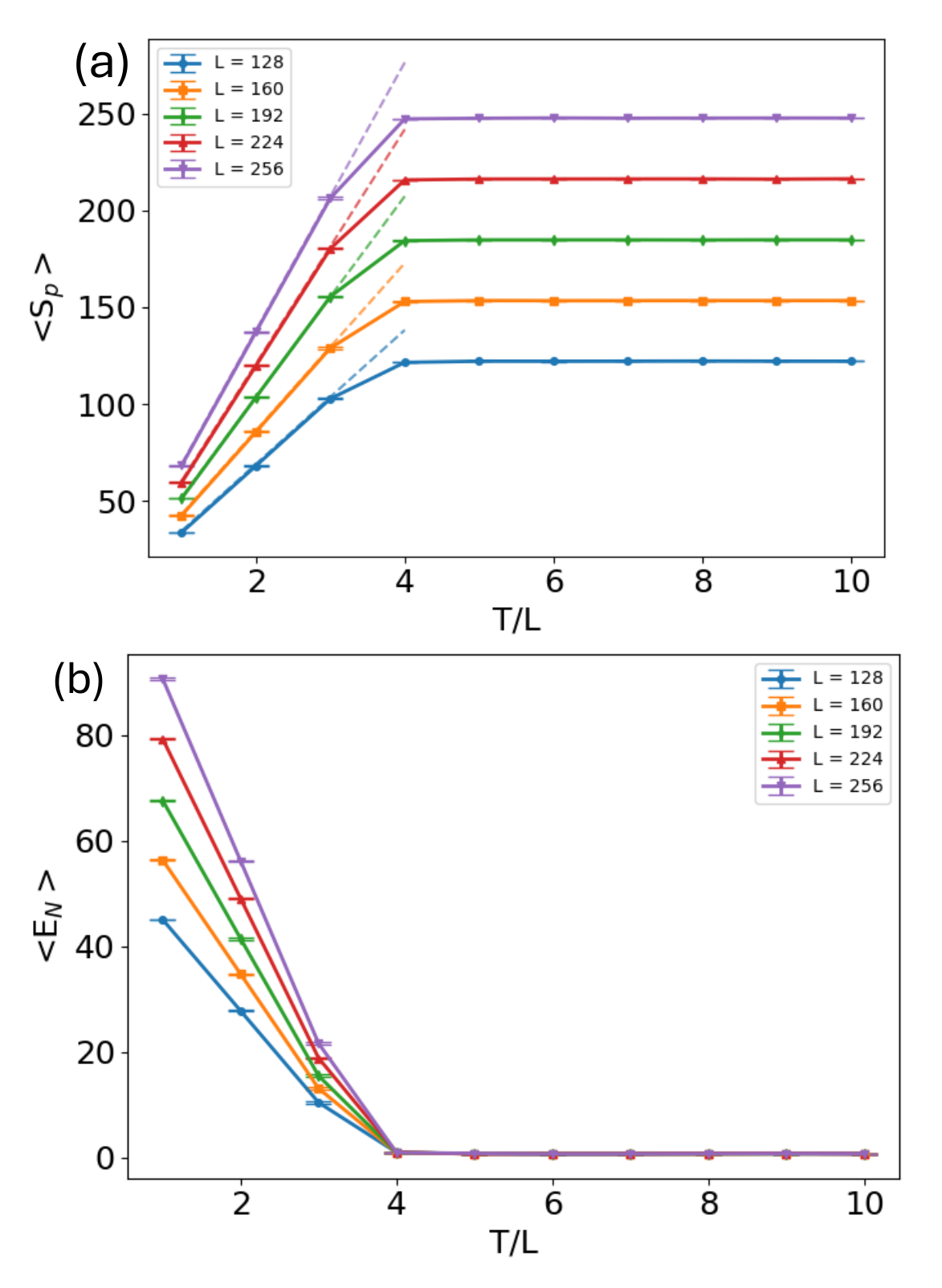}  
\end{center} 
\caption{
$T/L$-dependence of $S_p$ [(a)] and $E_N$ [(b)].We fix $p=0.25$. Both values saturate to a finite value at $T\sim 4L$ in which the reset-induced phase transition occurs. The data were averaged over $\mathcal{O}(10^3)$.}
\label{relaxation_time}
\end{figure}

We first show $T/L$-dependence for both $S_p$ and the MBN and with $p_c=0.25$ as shown in Fig.~\ref{relaxation_time} (a) and \ref{relaxation_time} (b). 
Both quantities reach a plateau around $T=4L$, indicating the phase transition occurs. 
In Fig.~\ref{relaxation_time} (a), the analytical result $S_p=Tp_c$ is plotted (dashed lines in Fig.~\ref{relaxation_time} (a)). We find that the slope of the numerical results matches with $p_cL$. In other words, the analytical results obtained in Appendix \ref{Appendix:B} also capture the numerical behaviors for the small-$T$regime. The analytical result ($Tp_c$) deviates from the numerical values near $T \sim 4L$. 
This also is understood by the same picture as the effective spin picture in the previous subsection. As shown in Appendix \ref{Appendix:B}, the small droplets of $\mathbb{I}$ emerge around the transition. It leads to the degeneracy of leading spin configurations and enhances the fluctuation. For larger $T/L$, the spin configuration with the domain wall becomes dominant, leading to the loss of $T/L$-dependence.

The MBN also exhibits a transition from the volume-law to the area-law phases at $T=4L$ as shown in Fig.~\ref{relaxation_time} (b). 
This behavior of the MBN exhibits the same correspondence to the one of $S_p$ as discussed in the previous subsection.

\begin{figure}[t]
\begin{center} 
\vspace{0.5cm}
\includegraphics[width=8.5cm]{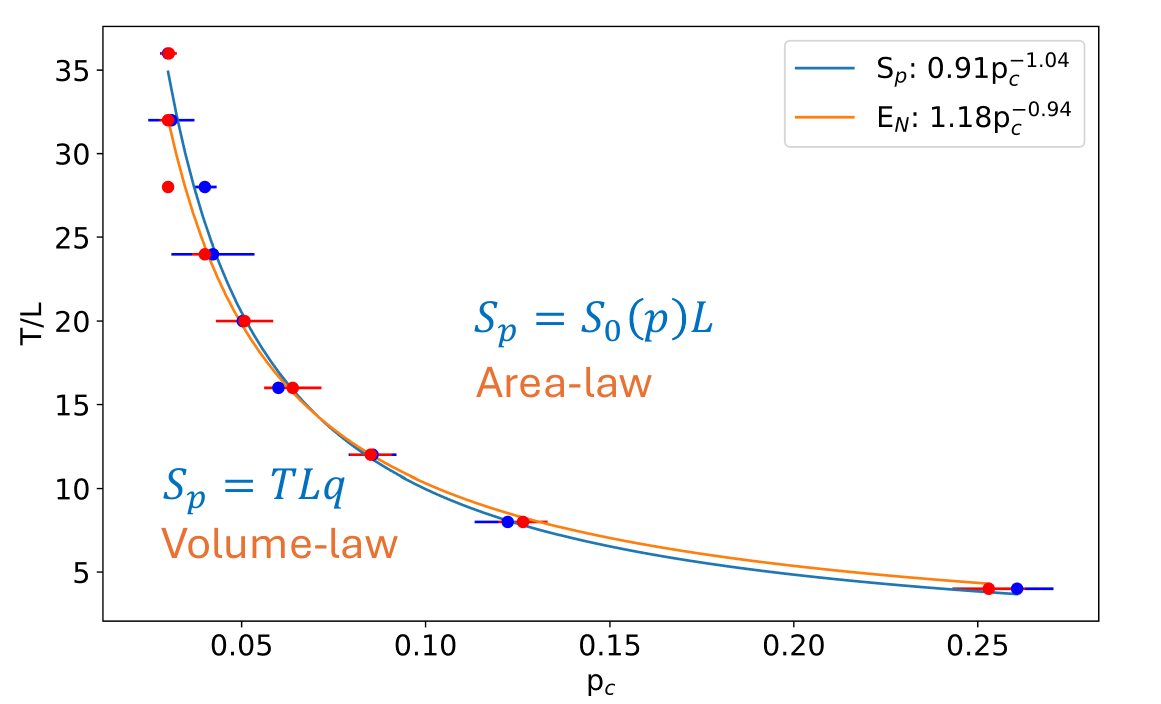} 
\end{center} 
\caption{
Phase-boundary curve of the reset-induced phase transition on $T/L$-$p$ plane. 
The curve is created by numerically estimated critical points $p_c$'s with estimated errors. The blue (orange) line is the estimated curve using $S_p$ ($E_N$). The regions above and below the curve correspond to distinct phases associated with each quantity. The averaged exponents over time interval from $4L$ to $36L$ on the phase boundary are $\nu=1.51(36)$ and $\zeta=0.49(18)$ for $S_p$ and $\nu=1.50(26)$ and $\zeta=0.43(19)$ for the MBN. The root-means-square-errors are computed. The results of critical exponents for each data set is in the main text.The fitting curve suggests that $T/L\propto1/p_c$. } 
\label{phase_transition_t:L>1}
\end{figure}

Next, we explore the reset-induced transition by varying $T/L$ and observing the MBN for various $p$ ranges. Here, the transition point is identified in the same way as the previous subsection. 
We summarize the critical transition points for $4\leq T/L \leq 40$ as shown in Fig.~\ref{phase_transition_t:L>1}, where the curve of the transition boundary is plotted in the $T/L-p$ plane. The blue (orange) line is represented as the estimated curve of the critical point obtained using $S_p$ (the MBN). The averaged critical exponents over all time steps are $\nu=1.50(26)$ for $E_N$ and $\nu= 1.51(36)$ for $S_p$. These values are close to $\nu\sim1.6$. Note that we perform the FSS analysis of $S_p$ in the form as follows:
\begin{equation}
    S_p(p)-S_p(p_c)=L^\frac{\zeta}{\nu}f(L^{\frac{1}{\nu}}(p-p_c)),
\end{equation}
because there is a significant size-dependent background contribution around the critical point seen in Fig.~\ref{second_Renyi_entropy}. This ansatz is conventionally used in Refs\cite{PhysRevLett.129.080501,PhysRevX.9.031009}. The results of critical exponents for each dataset are presented in Table \ref{table_2} and \ref{Table_3}. The reported values represent the critical exponents averaged over the time interval from $4L$ to $36L$, with the corresponding root-mean-square errors indicating their uncertainties. The area below (above) the curves indicates the volume (area) law phase from the viewpoint of the MBN. This also corresponds to the phase transition of $S_p$ indicated by Eq.~(\ref{s_p_transition}). These results suggest that $T/L$ is inversely proportional to $p_c$, $T/L \propto 1/p_c$. 
In other words, the reset-induced transition points highly depends on $T/L$. 
\begin{table}[h!]
\centering
\begin{tabular}{c|c|c}
\textbf{Time (L)} & \textbf{$\nu_{\mathrm{E_N}}$} & \textbf{$\zeta_{\mathrm{E_N}}$} \\
\hline
4  & $1.61\pm0.31$ & $0.43\pm0.17$\\
8  & $1.48\pm0.13$ & $0.54\pm0.19$\\
12 & $1.61\pm0.37$ & $0.57\pm0.33$\\
16 & $1.59\pm0.21$ & $0.55\pm0.22$\\
20 & $1.61\pm0.28$ & $0.59\pm0.28$\\
24 & $1.45\pm0.26$ & $0.21\pm0.04$\\
28 & $1.28\pm0.22$ & $0.35\pm0.03$\\
32 & $1.38\pm0.24$ & $0.22\pm0.08$\\
\end{tabular}
\caption{Values of $\nu_{\mathrm{E_N}}$ and $\zeta_{\mathrm{E_N}}$ for each time step from $4L$ to $32L$. The data for $36L$ are not reported due to the large statistical uncertainties associated with that time step.}
\label{table_2}
\end{table}
\begin{table}[h!] \centering \begin{tabular}{c|c|c} \textbf{Time (L)} & \textbf{$\nu_{\mathrm{S_p}}$} & \textbf{$\zeta_{\mathrm{S_p}}$} \\ \hline 4 & $1.73\pm0.27$ & $0.52\pm0.27$\\ 8 & $2.17\pm0.83$ & $1.66\pm0.27$\\ 12 & $1.29\pm0.09$ & $0.24\pm0.03$\\ 16 & $1.63\pm0.23$ & $0.71\pm0.19$\\ 20 & $1.28\pm0.40$ & $0.26\pm0.15$\\ 24 & $1.52\pm0.26$ & $0.49\pm0.30$\\ 28 & $1.43\pm0.21$ & $0.19\pm0.03$\\ 32 & $1.18\pm0.15$ & $0.21\pm0.08$\\ 36 & $1.37\pm0.25$ & $0.18\pm0.09$\\ \end{tabular} 
\caption{Values of $\nu_{\mathrm{S_p}}$ and $\zeta_{\mathrm{S_p}}$ for each time step from $4L$ to $36L$.}
\label{Table_3}
\end{table} 

We then comment on the two limit cases $p_c\rightarrow 0$ and $p_c\rightarrow L$. For $p_c\rightarrow 0$ ($q=0$), the system is in the volume-law because the unitary gates create more entanglement without intervention of the reset channels. Furthermore, for $p_c\rightarrow L$, ($q=1$), the system becomes a product state because the large amount of the reset channel interrupts the creation of entanglement. Our findings are consistent with the suggestion in \cite{Liu_2024}, except for the order of the phase transition. 

\section{Discussion}\label{discussion}
From the numerical results presented in Sec.\ref{numerical_result}, we give some discussions and interpretations as follows.

\subsection{$d$-dependence}
We map the random unitary quantum circuit onto the effective spin model. The analytical expression of $S_p$ is compared with the numerical data obtained. 
In the large-$d$ regime, a first-order phase transition is expected in \cite{Liu_2024}, characterized by a global change in the spin configuration (rather than spontaneous symmetry breaking). However, we observed numerically that for $d=2$ there is some deviation from the large-$d$ analytical results in Eq.~(\ref{s_p_transition}) around the critical point.
The reason of this deviation is that many-spin configurations that appear near the critical point ($p_c\sim0.25$ at $T=4L$) contribute to $S_p(d=2)$. 
The contributions of these spin configurations also lead to large fluctuations. From this point of view, in the numerical side in $d=2$, the sample-to-sample variance of $S_p$ shown in Fig.~\ref{second_Renyi_entropy} (b) increases near the transition point. The second derivative of $S_p$ shown in Fig.~\ref{second_order_phase_transition} (b) supports the fact that the reset-induced phase transition is continuous in the small $d$ regime. This observation is seen for the broad parameter regime $T/L$. As the critical point of the phase transition changes over time as shown in Fig.~\ref{phase_transition_t:L>1}, we expect that the phase transition is continuous for any arbitrary ratio $T/L$ in the small $d$ regime. 
However, detecting the threshold value of $d$ at which the transition changes from first-order (expected in \cite{Liu_2024,PRXQuantum.5.030327}) to second-order remains an interesting open question. We leave a detailed investigation of this question for future work.

\subsection{Monogamy relation}
In our work, we use the MBN to quantify quantum correlations within a mixed system rather than the bipartite entanglement entropy used in \cite{PhysRevX.10.041020,PhysRevX.9.031009}. Unlike the measurement induced phase transition, the reset channel induces the mixed phase transition. This originates from the initialization of qubit information by the reset channel. In the small-$p$ regime, we find that the purity $S_p$ increases linearly over time whereas the MBN decreases. This indicates that the system becomes increasingly more entangled with the environment rather than forming Bell's pairs within the system. This behavior is reminiscent of the monogamy of entanglement.

\subsection{Comment on critical exponent $\nu$}
We here comment the criticality for the reset-induced phase transition. We carry out the FSS analysis for various physical quantities. We find that the exponents $\nu$ for the variances of $S_p$ and $E_N$ using the data on the right side of the peak are much close to each other. In our system, we also note that the exponents $\nu$ extracted from the variances of $S_p$ and $E_N$ using the data in the region to the right of the peak are close to that obtained from the logarithmic negativity $\nu \sim 1.6$. This consistency suggests that the reset-induced transition may occur at a single point in $p$. Nevertheless, we refrain from drawing a definitive conclusion because of the limited accessible system sizes. In particular, the discrepancy between the exponents extracted from $E_N$ and $\mathrm{Var}(E_N)$ around the peak may arise from the strong sensitivity of $\mathrm{Var}(E_N)$ to sample-to-sample fluctuations near the critical point, analogous to susceptibility.

These numerical estimates through the exponents do not determine the universality class of the reset-induced transition. Although some exponents are close to known values (such as Ising CFT, $\nu=1$), this alone is insufficient to establish universality. Moreover, since the replica partition function in this case becomes unity in the replica limit, a CFT description would suggest a $c=0$ theory rather than the ordinary Ising CFT \cite{Jian_2020}. Thus, our FSS analysis provides evidence for a continuous transition with well-defined scaling, while the precise universality class remains an open question.

\subsection{Continuous phase transitions}
We have presented several numerical pieces of evidence supporting the conclusion that the reset-induced phase transitions are continuous. First, we observe the discrepancy of $S_p$ between the theoretical expectation from the effective spin model in the large-$d$ approximation and the numerical results in the $d=2$ case. This difference comes from the fluctuations due to the finite $d$ dimensional effects as expected in Ref.\cite{PRXQuantum.5.030327}. We claim that the fluctuations as quantified by the variances of $E_N$ and $S_p$ can significantly affect the order of phase transitions. Moreover, the second derivative of $E_N$ and $S_p$ tends to diverge near the critical point. The peak values of both observables increase as $L$ increases. Together with the FSS analysis, these results provide numerical evidence supporting the conclusion that the reset-induced phase transitions are continuous.

\section{Conclusion}\label{conclusion}
This work clarified detailed properties of the reset-induced phase transitions in a random unitary quantum circuit. 
The reset channel is feasible for the implementation of the real quantum circuits \cite{PhysRevLett.110.120501,Zhou_2021}. 
We focused on the reset-induced phase transition observed by $S_p$. The random unitary quantum circuit of the qubit ($d=2$) has strong fluctuations around the phase transition point, in contrast to the expectations obtained by the effective statistical model for large-$d$ \cite{Liu_2024,PRXQuantum.5.030327}. Our numerical data showed that the reset-induced phase transition in the qubit system is continuous. 
This indicates that the degree of qudit $d$ plays a significant role in determining the nature of the reset-induced phase transitions. 
We also verified that for the case $d=2$, the MBN exhibits the continuous transitions induced only by reset channels. Moreover, the relationship between the logarithmic purity $S_p$ and the MBN exhibits monogamy of entanglement as the strength of the reset channel approaches the reset-induced phase transition point.
In addition, we showed a shift of the critical point with $T/L$ inversely proportional to the critical reset probability $p_c$. We expect the same reset-induced phase transition to be observed for an arbitrary ratio of $T/L$. In particular, for the case $d=2$, 
our numerical results provide evidence for a continuous phase transition 
in contrast to the first-order behavior predicted by the large-$d$ analysis in the effective classical spin picture. 

There are several directions for future work beyond the scope of this study: (i) 
clarifying how the value of $d$ affects the critical exponent $\nu$ in random unitary quantum circuits with only reset operations and (ii) determining the corresponding conformal field theory for this reset-induced phase transition \cite{Zou2023}. It would be interesting to explore the physical phenomena associated with the reset-induced phase transition for future work.

\section*{ACKNOWLEDGMENTS}
This research was supported in part by JSPS KAKENHI Grant Numbers JP24K06878, JP24K00563, JP24K02948, JP21H04446, JP23K13026 and JP26K06956 and CSIS in Tohoku University and Akita University. Hinata Yokoyama acknowledges the support from GP-Spin at Tohoku University. The authors thank the Yukawa Institute for Theoretical Physics at Kyoto University. Discussions during the YITP workshop YITP-W-26-10 on "LOCALISATION 2026" were useful to complete this work.


\section*{Data availability}
The data that support the findings of this article are not publicly available. The data are available from the authors upon reasonable request.

\begin{appendix}
\numberwithin{equation}{section}

\renewcommand{\thesection}{A\arabic{section}} 
\renewcommand{\theequation}{A\arabic{equation}}
\renewcommand{\thefigure}{A\arabic{figure}}
\setcounter{equation}{0}
\setcounter{figure}{0}
\appendix

\setcounter{section}{0}
\renewcommand{\thesection}{\Alph{section}}
\makeatletter
\renewcommand{\theHsection}{\Alph{section}}
\makeatother



\makeatletter
\renewcommand{\theHfigure}{\thesection.\arabic{figure}}
\renewcommand{\theHtable}{\thesection.\arabic{table}}
\renewcommand{\theHequation}{\thesection.\arabic{equation}}
\makeatother
\section{Reset channel for two qubit example}\label{Appendix:A}
We here show how to act the reset channel for a two-qubit system described by a pure density matrix, $\rho=\ket{\psi}\bra{\psi}$ where $\ket{\psi}=\frac{1}{\sqrt{2}}(\ket{00}+\ket{11})$. 

Let us apply the channel to the first qubit. 
The trace part is expressed as $\rho_2=\mathrm{tr}_1(\rho)=\frac{\mathbf{I}}{2}$. Therefore, the state changes as
\begin{align*}
    \rho'=  R_{i=1}(\rho)= \ket{0}\bra{0}_1 \otimes \rho_2,= \ket{0}\bra{0}_1\otimes \frac{\mathbf{I}}{2}.
\end{align*}
This means that we have a state of $\ket{0}\bra{0}$ (product state) in the first qubit and a maximally mixed state in the second qubit. The state $\rho'$ is mixed. We next apply the channel on the second qubit. As a result, we obtaine the product state, $R_2(\rho')=\ket{00}\bra{00}$. 
This indicates that there is no entanglement between them.
Thus, for further many-qubit states, the reset channel induces both mixed and pure states. If we apply the channel to all sites, of course, we obtain simple product states. However, we do not focus on such a strong reset channel regime in terms of $q$.

\section{EFFECTIVE SPIN MODEL}\label{Appendix:B}
In this appendix, we briefly show the analytical expression for the second R\'{e}nyi entropy by mapping the circuit to the effective spin model. We make use of the analytical treatments suggested by \cite{Liu_2024}. 
In the mapping, instead of random Clifford gates, the Haar random unitary gate is assumed, also the degree of freedom of the physical bit (qudit) is introduced as $d$. We assume $d$ is large, i.e., $d\gg 1$.

By following \cite{Liu_2024,PhysRevB.101.104302,PhysRevB.108.L060302}, 
the replica logarithmic R\'{e}nyi-$n$ purity for the circuit is given by 
\begin{eqnarray}
S^{(n)}_{p}
=
\lim_{\substack{k\to 0}}
\frac{1}{k(1-n)}
\biggr[\log_2(Z_{up}^{(n,k)})-\log_2(Z_{0}^{(n,k)})\biggl],
\label{Sp}
\end{eqnarray}
where 
\begin{eqnarray}
Z_{up}^{(n,k)}
&=
\mathrm{Tr}
\biggr[
\left( C_{up}\right)^{\otimes k}
\left( \mathbb{E}_{u}\rho^{\otimes nk} \right)
\biggl]\nonumber
=
\mathrm{Tr}
\biggr[
\mathbb{C}_{up}
\left( \mathbb{E}_{u}\rho^{\otimes nk} \right)
\biggl],\label{Zup}\\
Z_{0}^{(n,k)}
&=
\mathrm{Tr}
\biggr[
\left( I_{up}\right)^{\otimes k}
\left( \mathbb{E}_{u}\rho^{\otimes nk} \right)
\biggl]\nonumber
=
\mathrm{Tr}
\biggr[
\mathbb{I}_{up}
\left( \mathbb{E}_{u}\rho^{\otimes nk} \right)
\biggl].
\label{Z0}
\end{eqnarray} 
$\rho$ is a mixed state and $\mathbb{E}_{u}$ represents the average manipulation for the Haar random unitary gates \cite{Liu_2024,PhysRevB.101.104302,PhysRevB.108.L060302,Vasseur2026}.  

In what follows we set $n=2$ ($S^{(2)}_p=S_p$). Then is $nk=2k$ and in Eqs.(A2) and (A3), $C_{up}$ and $I_{up}$ are given by
\begin{eqnarray}
C_{up} &=& \bigotimes_{i \in \mbox{all sites}} C_i, \\
I_{up} &=& \bigotimes_{i \in \mbox{all sites}} I_i,
\end{eqnarray}
where $C_i$ is a swap for two copies ($n=2$) qudit of the site $i$ and $I_i$ is the identity. 
$\mathbb{C}_{up}\equiv (C_{up})^{\otimes k}$ and $\mathbb{I}_{up}\equiv (I_{up})^{\otimes k}$ are permutation-operations associated to the real qudit. The label ``$up$'' means that permutation-operations are applied on qudits on the upper boundary of (1+1)-D space-time circuits. In what follows, we employ the notation of permutation-operations without  the label ``$up$'' if the operations are resided on the bulk of (1+1)-D space-time circuits.

After the replica trick shown above, 
the degree of freedom qudits $d$ in the replicated circuits is transformed by the permutation spins (corresponding to permutation-operations) denoted by $\sigma$. The element is the permutation group $\mathcal{S}_{2k}$\cite{PhysRevLett.129.080501,Liu_2024} and the total number of elements is $(2k)!$. Note that these spins $\sigma$ are not orthogonal \cite{PhysRevLett.129.080501,Liu_2024,Vasseur2026}. 
Especially, the identity $\mathbb{I}$ and the transposition $\mathbb{C}$ are important elements in $\mathcal{S}_{2k}$ in this work.

From the previous studies \cite{PhysRevB.101.104302,Liu_2024,PhysRevB.108.L060302}, $Z_{up(0)}^{(n=2,k)}$ is regarded as a partition function, the contributions of which are determined by the configurations of the classical permutation spins $\{\sigma\}$ on the 2D space (corresponding to the (1+1)-D space-time system). 
Now, $S_p$ of Eq.~(\ref{Sp}) is further written as
\begin{eqnarray}
S_{p}=
\lim_{k\to 0}
\biggr[-\frac{1}{k}
\biggr[\log_2(Z_{up}^{(n=2,k)})-\log_2(Z_{0}^{(n=2,k)})\biggl]\biggl]\nonumber
\equiv\lim_{k\to 0}
\biggr[-\frac{1}{k} \biggr[F^{(n=2,k)}_{up}-F^{(n=2,k)}_0\biggl]\biggl],
\label{Sp_def}
\end{eqnarray}
where $F^{(n=2,k)}_{up}\equiv -\log_2 Z^{(n=2,k)}_{up}$ and 
$F^{(n=2,k)}_0\equiv -\log_2 Z^{(n=2,k)}_{0}$. 
$F^{(n=2,k)}_{up(0)}$ is regarded as an effective free energy with the fixed boundary conditions represented by $\mathbb{C}_{up}$ and $\mathbb{I}_{up}$. 
Our interest here is to understand the qualitative behavior of $S_p$ by observing the free energies. In other words, our objective is to qualitatively estimate the effective partition function $Z^{(n=2,k)}_{up(0)}$ of $F^{(n=2,k)}_{up(0)}$. $Z^{(n=2,k)}_{up(0)}$ is given by Eqs.(\ref{Zup}) and (\ref{Z0}).

\begin{figure}[t]
\begin{center} 
\vspace{0.5cm}
\includegraphics[width=8.5cm]{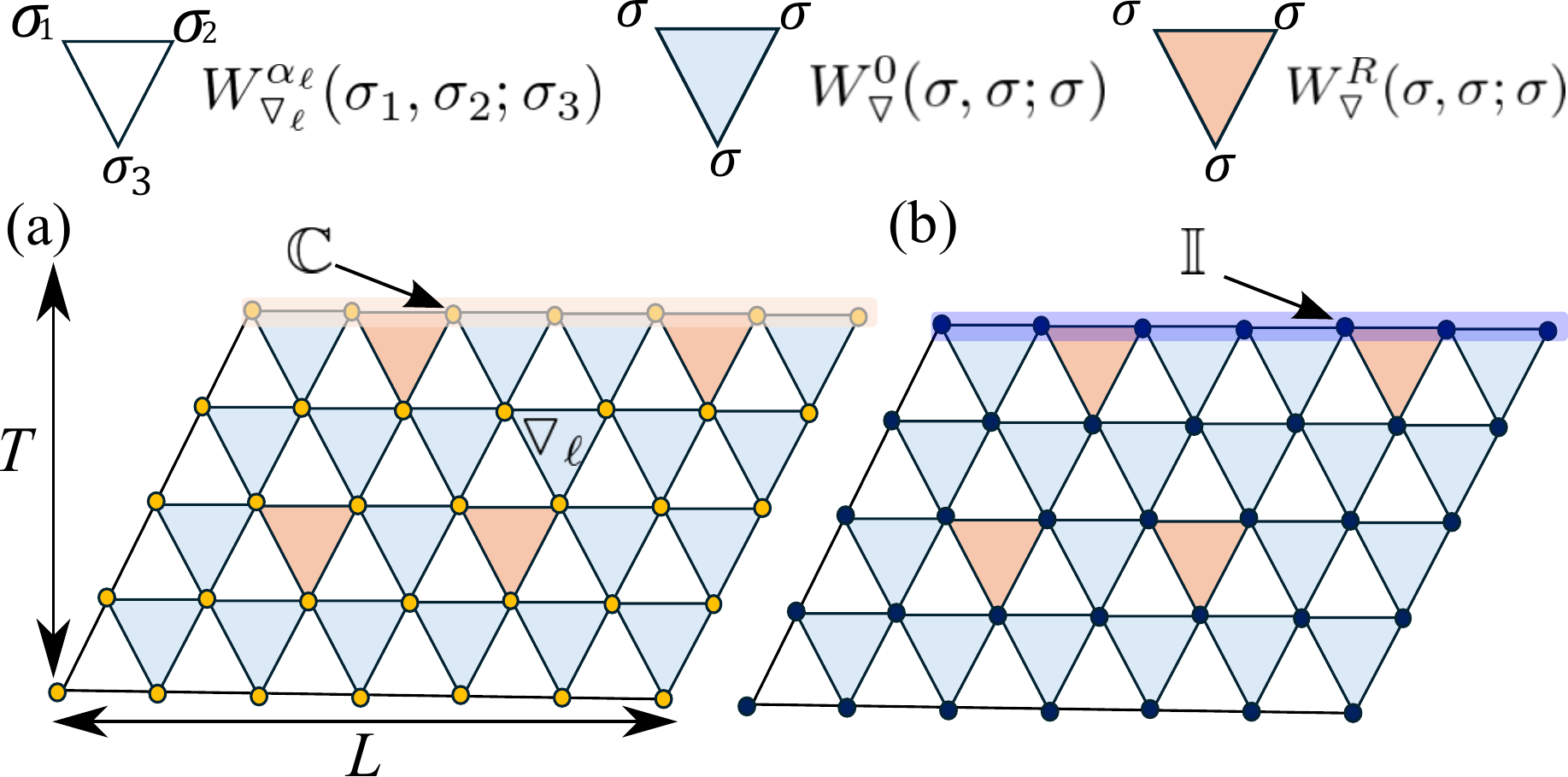}  
\end{center} 
\caption{(a) The leading contribution of the permutation spin configuration in $F^{(n=2,k)}_{up}$ (We assume $d$ is large).
The spin is fixed with $\mathbb{C}$ on the upper boundary and then the bulk ones are the same. (b) The leading contribution of the permutation spin configuration $F^{(n=2,k)}_0$. The spin is fixed with $\mathbb{I}$ on the upper boundary and then the bulk ones are the same. The orange (blue) colored sites represents $\mathbb{C}$ ($\mathbb{I}$).}  
\label{triangles_F_up_F_0}
\end{figure}

By following the previous works \cite{Liu_2024,PhysRevB.108.L060302,PhysRevB.101.104302}, our target random unitary quantum circuit under the reset channel can be mapped into 2D triangular lattice system where the classical permutation spin $\sigma$ is set on the site shown in Fig.~\ref{triangles_F_up_F_0}. 

The procedure of the mapping to the 2D triangular lattice system is following. 
Each Haar random unitary gates are averaged over \cite{Liu_2024,PhysRevB.108.L060302,PhysRevB.101.104302}, then the original two qudit lines are gathered by a single vertical line where the permutation spins are attached on the edges \cite{Liu_2024}. 
The connection to two-site random gates in the next time step is expressed by diagonal lines. The effects of the reset channel are imposed on the diagonal lines. Even without a reset channel, the diagonal lines give some contributions of the spin $\sigma$ to the partition functions \cite{Liu_2024,PhysRevB.108.L060302,PhysRevB.101.104302}. 
Then, the original circuit on (1+1)-D system becomes a honeycomb lattice \cite{PhysRevLett.129.080501,Liu_2024,PhysRevB.108.L060302}, where the spin $\sigma$ resides on the vertex. Furthermore, the vertex spins are effectively traced out to avoid a negative contribution to the partition function \cite{PhysRevLett.129.080501,Liu_2024}. As a result, the original circuit turns into a triangular lattice where the permutation spin $\sigma$ (such as $=\mathbb{I}$, $\mathbb{C}$) resides at the site(vertex). The schematic image is shown in Fig.~\ref{triangles_F_up_F_0}.

From the above manipulations, one reaches to the effective partition function $Z^{(n=2,k)}_{up(0)}$ which can be described by the configuration sum of the product of two types of triangle-weights determined by three spins ($\sigma_i$ with $i=1,2,3$) shown in Fig.~\ref{triangles_F_up_F_0}). The two types of the weight are denoted by $W^{\alpha_\ell}_{\triangledown_{\ell}}$ with $\alpha=\{0,R\}$. 
The partition function is written as
\begin{eqnarray}
Z^{(n=2,k)}_{up(0)}=\sum_{\{\sigma\}}\biggl[\prod_{\triangledown_{\ell}}W^{\alpha_\ell}_{\triangledown_{\ell}}(\sigma_1,\sigma_2;\sigma_3)\biggr],
\end{eqnarray}
where the label $\triangledown_{\ell}$ represents the position of the down-triangle weight $W^{\alpha_\ell}_{\triangledown_{\ell}}(\sigma_1,\sigma_2;\sigma_3)$ in the triangular lattice as shown in Fig.~\ref{triangles_F_up_F_0} (a). The triangle weight is determined by the three spins $\sigma$ on vertices of the down-triangle $\triangledown_{\ell}$ (The detailed form is discussed later). $W^{0}_{\triangledown_{\ell}}$ does not include the effect of the reset channel, while $W^{R}_{\triangledown_{\ell}}$ includes the effect of a single reset channel.

In the form of Eqs.~(\ref{Zup}) and (\ref{Z0}), $\mathbb{C}_{up}$ and $\mathbb{I}_{up}$ act as the upper-boundary conditions in the triangular lattice. 
That is, the spins on the upper boundary in the triangular lattice are fixed as $\mathbb{C}$ or $\mathbb{I}$ in Fig.~\ref{triangles_F_up_F_0}(a) and \ref{triangles_F_up_F_0}(b).
Thus, the estimation of $Z^{(n=2,k)}_{up(0)}$ corresponds to the estimation of the triangular weight $W^{\alpha_\ell}_{\triangledown_{\ell}}$ depending on the configuration of the permutation spin on the triangular lattice with the fixed boundary conditions. 

From now on, we move to the estimation of $Z^{(n=2,k)}_{up(0)}$ for both small- and large-$p$. In principle, it is difficult to treat the configuration sum $\sum_{\{\sigma\}}$, including the effect of the reset channels with probability $q=p/L$. We shall focus on the dominant spin configuration in the partition function $Z^{(n=2,k)}_{up(0)}$.
We emphasize that this approximation is suitable for the large-$d$ case.

\begin{figure*}[t]
\begin{center} 
\vspace{0.5cm}
\includegraphics[width=16cm]{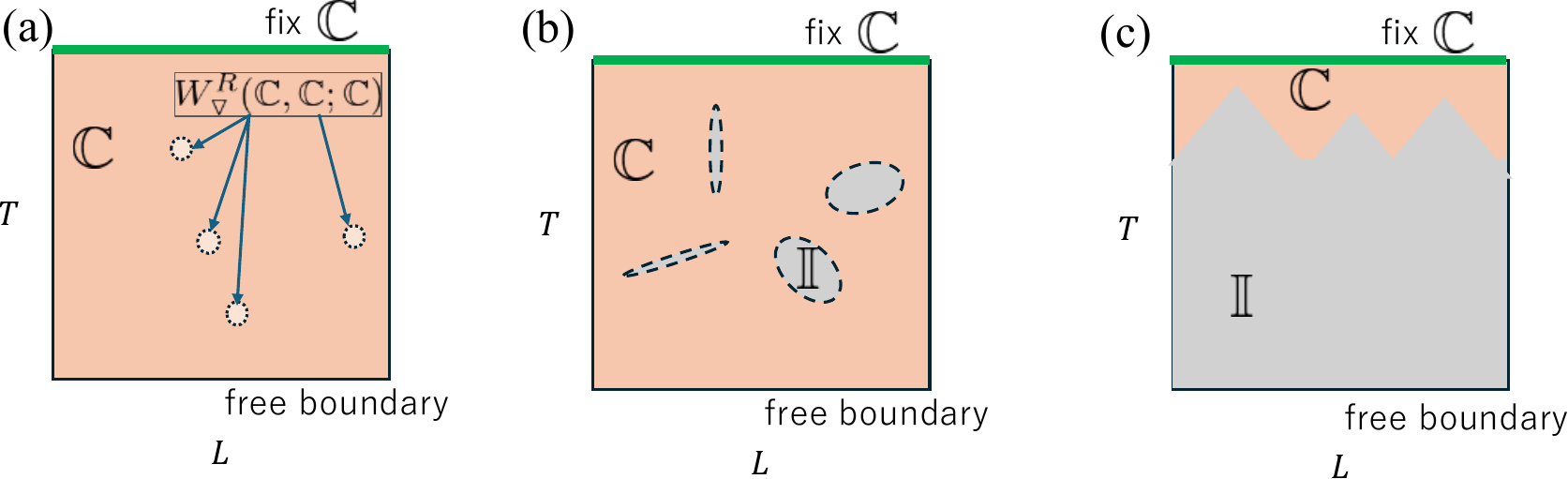}  
\end{center} 
\caption{Spin configurations becoming the leading contribution of partition function in $F^{(n=2,k)}_{up}$. (a) For $p<p_c$, all spin is set on $\mathbb{C}$ (b) Around the transition point $p\sim p_c$  local $\mathbb{I}$ droplets(bubbles) appear on the $\mathbb{C}$-background, where the leading spin configurations are degenerate and system-fluctuation gets large. (c) For $p>p_c$, the leading spin configuration is the simple two domain and there exists a single zig-zag crossing domain wall with the tension $S_0(p)$.}
\label{Fig3}
\end{figure*}

\subsection{Small-$p$ regime}
In the small-$p$ regime $p \ll 1$, we qualitatively estimate $S_p$. We consider the following assumptions:
\begin{enumerate}
\item We focus only on the spin configuration $\{\sigma\}$ giving a dominant contribution to the partition function. Indeed, we only pick up the leading contribution of $Z^{(n=2,k)}_{up(0)}$.

\item  We assume that the action of the Haar random gates effectively aligns the spins $\sigma$ on the triangular lattice depending on the fixed spins at the upper boundary, so that the dominant spin configuration is uniform.

\item We assume that reset channels are dilute on the triangular lattice and primarily affect the diagonal lines as in \cite{Liu_2024}.
We further assume that adjacent diagonal lines do not simultaneously host reset channels.
In this picture, the reset-channel contribution effectively reduces to a single diagonal line to the triangular-weight $W^{\alpha_\ell}_{\triangledown_{\ell}}$. The corresponding triangular weights denoted by $W^{R}_{\triangledown}(\sigma,\sigma;\sigma)$ are assumed to be sparsely distributed over the lattice as shown in Fig.~\ref{triangles_F_up_F_0}.
\end{enumerate}
Based on these assumptions, we express the leading terms for $Z^{(n=2,k)}_{up}$ and $Z^{(n=2,k)}_0$. We estimate the contribution for $d\gg 2$. 
The two triangular weights are known to become \cite{Liu_2024}:
\begin{eqnarray}
W^{R}_{\triangledown}(\sigma,\sigma;\sigma)\sim d^{-|\sigma|},
\end{eqnarray}
where $|\sigma|$ is the number of transpositions $\sigma$ from the identity permutation spin $\mathbb{I}$ 
and
\begin{eqnarray}
W^{0}_{\triangledown}(\sigma,\sigma;\sigma)\sim d^{0}=1.
\end{eqnarray}
By using these weights, 
the leading of the configuration of $Z^{(n=2,k)}_{up}$ denoted by $Z^{l}_{up}$ is the all-spin-$\mathbb{C}$ configuration due to the spin $\mathbb{C}$ on the upper boundary shown in Fig.\ref{Fig3}(a): 
\begin{eqnarray}
Z^{l}_{up}=\biggl[W^{R}_{\triangledown}(\mathbb{C},\mathbb{C};\mathbb{C})\biggr]^{TLq}\biggl[W^{0}_{\triangledown}(\mathbb{C},\mathbb{C};\mathbb{C})\biggr]^{N_0}\nonumber
\sim \biggl[W^{R}_{\triangledown}(\mathbb{C},\mathbb{C};\mathbb{C})\biggr]^{TLq},
\end{eqnarray}
where $N_0=(1-q)TL$. 
Thus, we estimate the free energy as
\begin{eqnarray}
F^{(n=2,k)}_{up}\sim-\log_2 [Z^{l}_{up}]=-\log_2[(d)^{-kTLq}].
\label{Fup_psmall}
\end{eqnarray}
where we used $|\mathbb{C}|=k$.

The reason why we set all $\mathbb{C}$ is as follows. If we locally employ $W^{R}_{\triangledown}(\mathbb{I},\mathbb{I};\mathbb{I})$ under the presence of reset channel, surely $W^R_{\triangledown}=1$, but this is not optimal configuration since the change of spin $\mathbb{I}$ induces the domain wall (DW) to the bulk spin $\mathbb{C}$. As a result, totally the DW decreases the contribution of $Z_{up}$. Thus, at least, the local change $W^{R}_{\triangledown}(\mathbb{C},\mathbb{C};\mathbb{C})\longrightarrow W^{R}_{\triangledown}(\mathbb{I},\mathbb{I};\mathbb{I})$ is not suitable. On the other hand, 
the leading term of the configuration of $Z^{(n=2,k)}_{0}$ denoted by $Z^{l}_{0}$ is all-spin-$\mathbb{I}$ due to the spin $\mathbb{I}$ on the upper boundary. We obtain $Z^{l}_{0}=1$, which leads to $F^{(n=2,k)}_{0}=0$. 

From these considerations, the logarithmic purity $S_p$ is estimated as
\begin{equation}
S_{p}
=\lim_{k\to 0}\biggr[\frac{1}{k}(kTLq)\biggl] \log_2 (d)=Tp\log_2(d). 
\end{equation}
Note that the replica limit gives no effects to the logarithmic purity.

Here, we further set $d=2$ as an example, corresponding to the case consistent with our numerics. Then, we obtain 
\begin{equation}
S_{p}=Tp.
\label{Sp_analytical}
\end{equation}
As shown in the main text, compared to the data of Fig.~\ref{second_Renyi_entropy} in which $T=4L$, $p$ and $L$ dependencies shows strong correspondence in the small-$p$ regime. The analytical estimation captures the numerical behaviors.

Here, we also give a phenomenological image to the behavior $S_p$. 
In general we expect that the behavior of $S_p$ in $d=2$ case deviates from the behavior of the form of Eq.~(\ref{s_p_transition})
, especially around the phase transition. There, we expect that in the spin mapping, due to the presence of reset channels, small $\mathbb{I}$ bubbles and short DWs are formed as shown in Fig.\ref{Fig3}(b). 
In particular, the assumption of picking up the dominant spin configuration to estimate $Z^{l}_{up}$ becomes less accurate around the transition point and we expect that the fluctuation in the system gets large. This expectation appears in our numerical results in Fig.~\ref{second_Renyi_entropy}.





\subsection{Large-$p$ regime}
We next discuss the large-$p$ regime. This regime has been mainly discussed in \cite{Liu_2024}. 

Even for the upper boundary with the spin $\mathbb{C}$ fixed, the spin $\mathbb{I}$ proliferates due to the large-$p$ in the background $\mathbb{C}$. As $p$ increases, domain $\mathbb{I}$ increases. As a result, two large domains appear for $\mathbb{I}$ and $\mathbb{C}$ as shown in Fig.~\ref{Fig3}(c), where the domain $\mathbb{C}$ attaches to the upper boundary. Here, the DW is zig-zag, separating the domains \cite{Liu_2024}. The leading contribution of $Z^l_{up}$ comes from the contribution $W^{0}_{\triangledown}(\mathbb{I},\mathbb{C};\mathbb{I})$,  $W^{0}_{\triangledown}(\mathbb{C},\mathbb{I};\mathbb{I})$, which occurs on the DW path. 
Here, as for \cite{Liu_2024}, since $W^{0}_{\triangledown}(\mathbb{I},\mathbb{C};\mathbb{I})=W^{0}_{\triangledown}(\mathbb{C},\mathbb{I};\mathbb{I})$, the leading term is approximately
\begin{eqnarray}
Z^{l}_{up}&\sim&\biggl[W^{0}_{\triangledown}(\mathbb{C},\mathbb{I};\mathbb{I})\biggr]^{L_{DW}}
\sim (d^{-|\mathbb{C}^{-1}\mathbb{I}|})^{L_{DW}},
\end{eqnarray}
where $L_{DW}$ is the domain-wall length, approximately equal to $L_{DW}=S_0(p)L$ \cite{PhysRevLett.129.080501,Liu_2024} where $S_0(p)$ is an effective tension.
Moreover, we use $|\mathbb{C}^{-1}\mathbb{I}|=k$. 
It leads to 
\begin{eqnarray}
F^{(n=2,k)}_{up}\sim-\log_2 [Z^{l}_{up}]=kS_0(p)L\log_2 d.
\label{Fup_psmall}
\end{eqnarray}
Thus, we obtain 
\begin{equation}
S_{p}
=S_0(p)L\log_2 d.
\end{equation} 
Phenomenologically, in the large-$p$ regime the logarithmic purity $S_{p}$ comes from the DW energy as a string tension form $S_0(p)L\log_2 d$. 
In addition, we expect that $S_0(p)$ does not depend on the system size $L$. This also can be regarded as the appearance of the volume-law scaling of the logarithmic purity.
Again, we take $d=2$. Then, $S_p = S_0(p)L$. This form agrees with the behavior of $S_p$ observed in our numerical calculations.\\ 

\subsection{Intermediate $p$ regime}
We additionally comment on the form of $S_p$ in Eq.~(\ref{s_p_transition}) 
in the intermediate regime where $p\sim p_c$.
In the small-$T$ regime, the numerical results in the main text show strong agreement with the analytical expression of $S_p$ given in Eq.~(\ref{s_p_transition}) as shown in Fig.~\ref{relaxation_time} (a). However, around the transition value of $T$, a discrepancy between the analytical and numerical results is observed. This is because the fluctuations in the system increase as small $\mathbb{I}$ bubbles and short DWs are formed as shown in Fig.\ref{Fig3}(b) in the effective spin picture. In other words, the leading spin configurations in the intermediate $p$ regime are degenerate, which significantly enhances the logarithmic purity. Consequently, the analytical expression for $S_p$ deviates from the numerical results in this regime.

\section{Large-$q$ limit }\label{Appendix:C}
\begin{figure}[t]
\begin{center} 
\vspace{0.5cm}
\includegraphics[width=8.5cm]{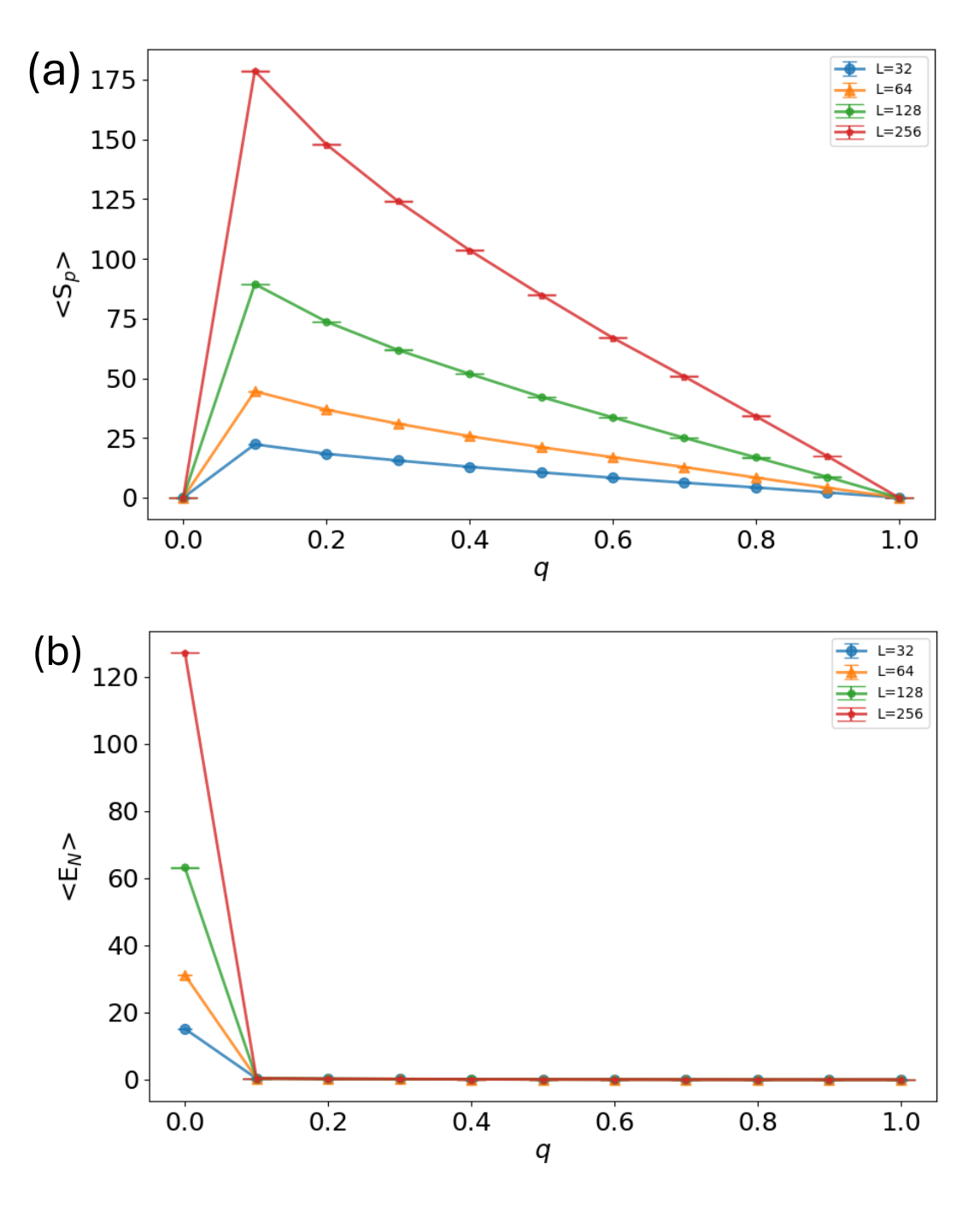}  
\end{center} 
\caption{Broad $q$-dependence of a reset channel in the random Clifford circuit. (a) $q$-dependence of $S_p$. (b) $q$-dependence of $E_N$. All the data were averaged over $\mathcal{O}(10^3)$ trajectories and we set $T=4L$.}  
\label{the_role_of_reset_channel}
\end{figure}

We show the effect of the reset channel in the large-$q$ case. The channel tends to lead to the pure and product states if the application of the channel is spatially broad, corresponding to taking the large $q$ value in our setup. We verity this trend by observing both $S_p$ and $E_N$. 

Fig.~\ref{the_role_of_reset_channel}(a) show $q$-dependence of $S_p$. With $q=0$, the system is maximally entangled in the late time $T=\mathcal{O}(L)$ and the state is pure. With an increase of $q$, the system starts to be more mixed due to the random unitary gates and reaches the maximum value with $q\sim 0.1$. Then, the system becomes purified with $q=1$ (product state), and the system-size dependence vanishes. Thus, the large reset channel induces purification even for many-qubits case in the presence of random unitary gates. Moreover, the MBN is plotted as shown in Fig.~\ref{the_role_of_reset_channel}(b). When the system is maximally entangled with $q=0$, we observe that the reset channel suddenly breaks the quantum entanglement in the system shown as a sudden drop around $q=0.1$.

From these observations, in $0\leq q\le0.1$, specifically $q\sim 0.1$, a reset channel makes a system more mixed, reducing the entropy of internal entanglement. As $q$ increases, the system becomes purified and the internal entanglement vanishes. Finally, in the limit $q\to 1$, the channel is applied to all sites, and the state becomes $|0\rangle^{\otimes L}$ due to projection in the reset channel. However, we emphasize that this sudden change in $S_p$ and the MBN corresponds to the phase transition at $T=4L$ seen in the subsection A in Sec.~\ref{numerical_result}. Because the interval we take for Fig.~\ref{the_role_of_reset_channel} is considerably larger than the one in the subsection, we simply could not capture the phase transition in the large $q$ case (recall $q=p/L$).

\section{The FSS analysis of Variances around the peaks}\label{variance_appendix}
\begin{figure}[t]
\begin{center} 
\vspace{0.5cm}
\includegraphics[width=14cm]{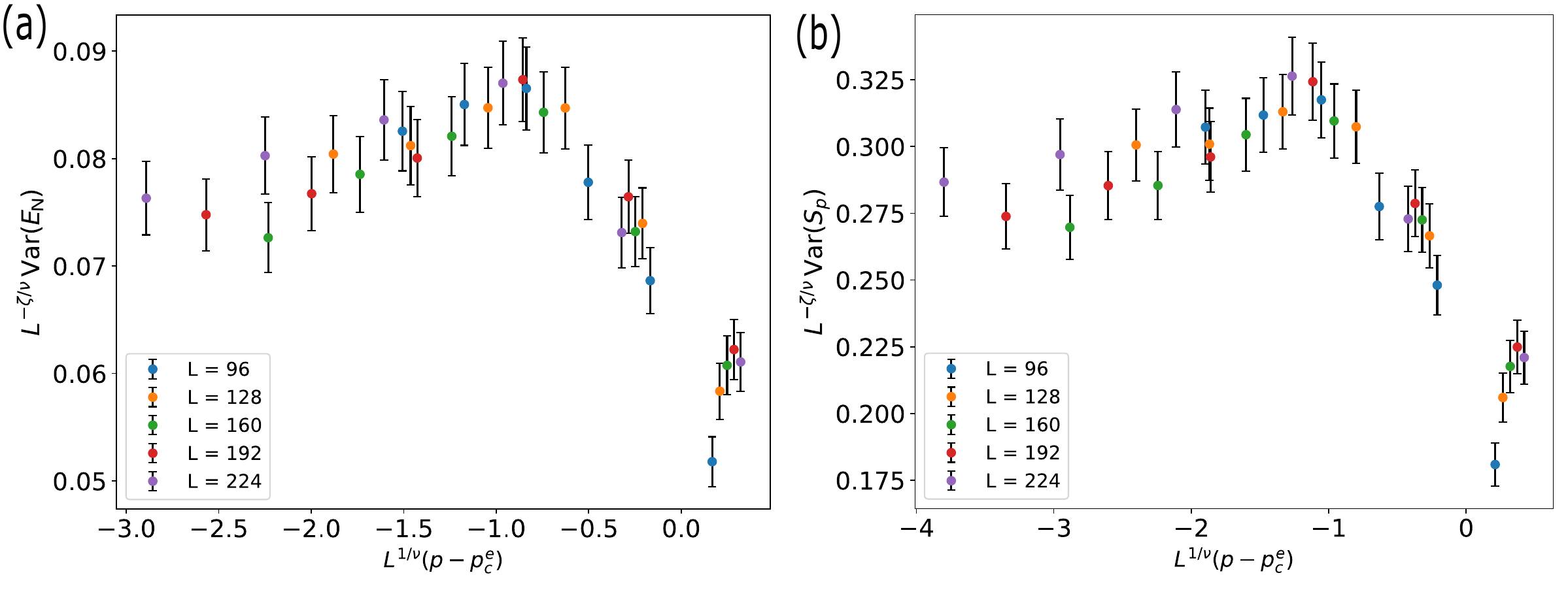}  
\end{center} 
\caption{
The results of the FSS analysis of $\mathrm{Var}(E_N)$ [(a)] and $\mathrm{Var}(S_p)$ [(b)] around the variance peaks. The results are severely affected by finite-size effects.}
\label{variance_peak_fss}
\end{figure}
The result of the FSS analysis of $\mathrm{Var}(S_p)$ and $\mathrm{Var}(E_N)$ around the peak is shown in Fig.\ref{variance_peak_fss}. Here, we use $p_c^e$, the effective critical point, and followed the ansatz of Eq.\ref{new_ansatz_S_p} in the main text.
While the quality of the data collapse is limited by the relatively large uncertainties and finite-size effects, we show this result as an additional FSS trial.
For $\mathrm{Var}(E_N)$, the extracted parameters shown in Fig.~\ref{variance_peak_fss}(a) are $p_c^e=0.23$, $\nu=1.30(19)$, and $\zeta=1.53(12)$. For $\mathrm{Var}(S_p)$, shown in Fig.~\ref{variance_peak_fss}(b), we obtain $p_c^e=0.23(14)$, $\nu=1.22(44)$, and $\zeta=1.44(34)$.
However, these results might not accurately capture the critical behavior due to the relatively large uncertainties.
The values, obtained from the data near the peak, $\nu_{S_p}\sim 1.2(44)$ and $\nu_{E_N}\simeq1.3(19)$ are strongly affected by severe finite-size effects. We note that these estimates are inconsistent with either Ising CFT or 2D percolation universality class, regarded as possible universality classes to which MIPTs may belong \cite{PhysRevX.10.041020,Yaodong_2018,Yaodong_2018,Zabalo_2020,Jian_2020,Bao_2020,PhysRevX.9.031009,Fisher_2023}.

\end{appendix}





\bibliography{SciPost_Example_BiBTeX_File.bib}


\end{document}